# A multimodal solution approach for mitigating the impact of planned maintenance on metro rail attractiveness

A. Consilvio[1], L. Calabrò, A. Di Febbraro and N. Sacco

*Abstract*—The possible unavailability of urban rail-based transport services due to planned maintenance activities may have significant consequences on the perceived quality of service, thus affecting railway attractiveness.

To cope with the mitigation of planned service interruptions and to guarantee a seamless journey and a good travel experience for passengers, it is possible to exploit the existing services differently and/or provide additional on-demand services, such as temporary supplemental bus lines.

In this context, this paper aims to develop a mathematical programming model for planning service interruptions due to maintenance considering passenger transport demand dynamics. In particular, the proposed approach deals with service interruptions characterized by a long duration for which timetable adaption strategies are not applicable, suggesting mitigation actions that exploit the already existing services and/or the activation of additional ones, with the aim of minimizing users' inconvenience. In doing so, the planned infrastructure status (i.e., available or under maintenance), as well as the forecasted transport demand, are taken into account to adapt the service accordingly by offering a multimodal transport solution to passengers.

To find the best solution, a decomposition solution approach is proposed in combination with a multistage cooperative framework with feedback that models the negotiation process between the involved actors.

Finally, the applicability of the proposed approach to real case studies is discussed based on some performance indicators.

I. INTRODUCTION

In the rail sector, performing preventive maintenance tasks during a time slot normally requires possession of the track, which implies a complete capacity breakdown of the service since no trains are allowed to run during such a possession.

Therefore, the interaction between train circulation and maintenance interventions may be critical; in fact, when train-free intervals at night are insufficient to perform maintenance activities, adapting and modifying train timetables cannot be a sufficient mitigation strategy. Such a problem is worsened by the fact that rail transport is usually characterized by few path alternatives, and integrated management of the multimodal transport network is necessary to guarantee the achievement of an acceptable level of service availability and reliability, mitigating the impact of infrastructure possessions.

This is particularly convenient in urban areas, where there are many alternative solutions, such as metro lines and public ground transport services (i.e., light trains and buses). In this framework, either in the case of a single multimodal public transport provider or in the case of different providers, cooperation among the different modes turns out to be a key aspect for achieving common goals of efficiency, reliability, and resiliency. In contrast, in extraurban scenarios, where few alternatives exist, the rail service provider is called to organize and plan additional on-demand services in case of planned maintenance and infrastructure possession.

The main innovation introduced by this paper consists of the definition of a methodology that enables infrastructure managers and service operators to find a good service interruption planning and a set of mitigation actions, considering both the integration with existing lines and the introduction of ad hoc services. In particular, this paper aims to develop a mathematical programming model with two goals:
1. planning infrastructure possessions for maintenance, dealing with interventions that require the interruption of train circulation for the whole day during a nonnegligible period of time, such as a week;
2. planning mitigation actions, such as alternative services, since in this situation, the adoption of timetable replanning strategies is not possible because rail services are suspended and alternative transport solutions must be provided to passengers.

The proposed approach suggests a suitable maintenance schedule and some mitigation actions capable of reducing the impact of rail service interruptions by means of integration with existing alternative services and/or the activation and planning of additional ad hoc bus services, with the aim of minimizing users' inconvenience and service failure risk. It is worth emphasizing that, in doing so, the approach considers *tactical planning* of infrastructure possessions for maintenance and mitigation decisions

A. Consilvio, A. Di Febbraro, and N. Sacco are with the Department of Mechanical, Energy, Management and Transportation Engineering (DIME), University of Genoa, Italy (corresponding author: alice.consilvio@unige.it). L. Calabrò is with Hitachi Rail STS.

that can be made in advance, having sufficient time to organize the multimodal service and to inform the users. The considered approach is therefore an offline planning model that does not consider unplanned real-time service interruptions.

Regarding the solution approach, a multistage cooperative framework with feedback that mimics the usual real-world decision-making is proposed to represent the negotiation process between different actors:
1. the infrastructure manager;
2. the metro service provider;
3. service providers of other modes.

According to this scheme, the general problem is divided into three subproblems that represent different decision moments:
1. the booking of maintenance possessions (subproblem SP1);
2. the decision to use existing services or activate ad hoc solutions (subproblem SP2);
3. the design of the alternative proposed solution (subproblem SP3).

The proposed approach will allow public transport operators to better allocate their resources, resulting in lowering the costs of operations and increasing service availability. This will reinforce the capacity to better prepare for events (the temporary interruption of a line section), improving preventive planning and resource allocation. At the same time, the planned infrastructure status (i.e., available or under maintenance), as well as the forecast transport demand, are exploited to adapt the service accordingly and to offer a resilient multimodal transport solution to passengers. Finally, the information regarding rail service interruptions and the proposed alternative solutions can be communicated to passengers in advance, thus reducing the negative impact on users' travel experience.

It is worth noting that the considered solution scheme enables the capability of considering different implementations of the three subproblems, such as, for instance, the possibility of changing subproblem SP1 to introduce new details or a particular customization, leaving the others unchanged. In this connection, the generality of the solution algorithm is guaranteed if the kind of inputs and outputs of the three problems remains the same.

To demonstrate the practical significance of the proposed method and discuss the computational effort, two case studies based on the public transport networks of the Italian cities of Genoa and Milan are presented, showing that the joint management of planned interruptions can significantly reduce the cost for service providers and the impact on travelers' experience.

## II. LITERATURE REVIEW

To meet the growing passengers' transport demand and achieve higher competitiveness in a multimodal transport system, the infrastructure needs to be well utilized (in terms of transport service). Meanwhile, railway infrastructure should be well maintained by means of preventive maintenance. With the development of preventive and predictive maintenance strategies, studies that deal with policies for planning inspections, maintenance activities and renewals according to asset status have been developed [1], [2]. The aim is to prevent failures and increase system reliability [3], [4], [5], [6]. These studies consider predefined maintenance windows with no concern about how the interaction with train operations should be resolved.

Previous studies on interactions with train operations are mainly based on two different approaches: scheduling possession windows by minimizing the maintenance impact on the service [7], [8] or vice versa dealing with adjusting timetables according to maintenance possessions [9], [10], [11].

Liden et al. [7] and Kalinowski et al. [8] face the problem of planning maintenance access windows considering train timetables, developing models for the assessment and dimensioning of maintenance windows, and allocating train-free slots reserved for maintenance tasks on railway infrastructure.

In Van Aken et al. [9], [10] the train timetable adjustment problem (TTAP) is addressed. Given an existing timetable with scheduled arrival, departure and passing times and a list of unavailable track sections due to maintenance, the goal is to automatically generate alternative railway timetables. The objective is to minimize the number of cancellations and the deviation from the originally scheduled event times in the timetable.

Zhang et al. [11] present an optimization problem for integrating retimetabling and rerouting decisions within station areas for multiple stations when scheduled maintenance makes the existing, optimized train schedules infeasible.

Currently, coordinated, and integrated approaches for planning infrastructure maintenance and train operations are receiving greater attention. The objective is to reduce delays for both trains and maintenance works. Forsgren et al. [12], D'Ariano et al. [13] and Luan et al. [14] simultaneously optimize trains' routes, orders, departure times and arrival times at passing stations, as well as the work time of maintenance tasks within an integrated approach.

Studies in the literature can be divided into online and offline problems. Online problems deal with unexpected maintenance disturbances [15], [16], [17], while offline models address the topic of planning maintenance activities in advance [18], [19].

Albrecht et al. [15] develop an operational tool to quickly generate a revised schedule taking into account the new state of a disrupted system. Arenas et al. [16] consider unplanned maintenance activities that have to be introduced at short notice and propose a mixed-integer linear programming formulation that rearranges train timetables with respect to the new unavailabilities.

Zhang et al. [17] address track emergency maintenance of high-speed railways by presenting a rolling horizon framework realizing online feedback correction and improving the robustness of the control strategy in a dynamic environment in which disruptions occur.

Zhang et al. [18] integrate the train timetabling problem and track maintenance task scheduling considering two types of maintenance activities: (1) ordinary maintenance activities, which require a relatively short time to execute and are often scheduled during a minor possession, and (2) major maintenance, which involves renewing or repairing the tracks and reduces the available railway capacity significantly, since no train can occupy the maintenance tracks during the long work time. The goal is to achieve an effective trade-off between the minimization of the total train travel time and the maintenance tardiness cost.

Zhang et al. [19] formulate a joint optimization model for train scheduling and maintenance planning problems to minimize the total travel time of passengers and maintenance costs simultaneously. A layered space-time network model with macroscopic geography and microscopic discrete time representation is applied. The results lead to a high-quality timetable for passengers and a low maintenance cost for railway departments.

All these approaches try to solve the issue of service interruption due to maintenance considering possible solutions within the rail system environment.

Nevertheless, the strategy of adapting and modifying trains' timetables is often not possible due to insufficient rail line capacity and the extended time required to execute maintenance activities, implying the closure of a line section for many days [20].

In this context, cooperation with other transport modes has become key to achieving a high level of service availability.

The present paper addresses this issue of enhancing cooperation between transport operators by proposing a cooperative framework to model the negotiation process between the different actors involved.

The aim is to cover an aspect that is usually neglected in existing studies on maintenance window planning and service disruption management.

Existing studies usually deal with the problem of service interruption from the travelers' point of view, developing trip planning models capable of suggesting the best path to passengers in case of service downtimes [21].

Nevertheless, the problem of integrated management of planned interruptions from the point of view of service providers deserves particular attention to reduce the impact on rail service attractiveness.

Existing works usually distinguish between unexpected and planned service disruptions.

Planned disruptions differ from unplanned disruptions in many ways. A first 'chaotic' phase is observed immediately after an unplanned disruption. During this phase, operators and passengers have limited or no visibility of the cause, magnitude, or duration of the failure. This phase does not exist in the case of planned works, especially when communication campaigns are in place. When unexpected incidents occur, an important segment of the demand is obliged to make in-trip rerouting and real-time decisions with limited information at hand, whereas during planned disturbances, the route is decided in advance. Information availability changes for the operators as well. Furthermore, the time horizon is entirely different, so system equilibrium may be possible for planned works but not for short disruptions. Additionally, planned disturbances allow operators to predict and ensure the availability of necessary resources, both human and technical. In conclusion, the best strategies and disruption recovery problems may differ significantly from one case to the other.

Regarding unexpected disruptions, studies of contingency planning are presented in the literature [22]. In particular, the bus bridging problem is considered for unexpected disruptions that affect metro lines. The term 'bus bridging' refers to a temporary ad hoc service that restores the connectivity between affected stations to reduce the impacts of metro disruptions [23]. They consider unexpected disruptions and the goal of reducing passengers' inconvenience [24]. In such instances, quick and efficient substitution of services is necessary for accommodating metro passengers. These studies are also known as 'passenger evacuation bus bridging'[25], [26].

Despite the numerous studies on unplanned disruptions, few investigate planned disruptions [27].

In [28], a mathematical model is presented that aims at achieving the maximum synchronization of two public transport systems, while in [29], the goal of coordinating the bus transit network is achieved by determining the bus line alignment and operating frequency that can minimize the total passenger and operator costs in the public transit system.

To increase the resiliency of metro services, studies on planned service interruptions deal with possible integration with existing bus services instead of designing new bus lines [30]. For example, they change the path of some existing lines to minimize the impact of a service interruption. Some works in the literature take into account different available transport modes in a cooperative planning framework [31], [32].

Nevertheless, the collaborative joint management of planned service interruptions has not yet received the attention that it deserves, and the chain of decisions required by this problem is still managed manually by service providers [33]. Therefore, a model for mitigating the impact of infrastructure possession could significantly speed up this process and guarantee the design of more efficient solutions, leading to higher service quality and greater passenger satisfaction.

In this paper, a novel model is proposed that enables operators to evaluate different management schemes of planned service interruptions, considering both the integration with existing lines and the introduction of ad hoc services. Moreover, the planning of maintenance windows and the management of service interruptions are considered in a unique framework, with a twofold objective: planning the possessions minimizing the impact on passengers and providing to passengers the best service considering the integration with other transport modes.

Compared to the reference literature, the contributions of this paper consist of:
- planning maintenance by taking into account the impact on users and rail transport attractiveness, considering the main goals of both the infrastructure manager and the metro service operator. In particular, compared to [1-6], the presented approach solves the problem of booking maintenance windows considering also the interaction with train circulation; while compared to [7-14], it refines the search for an optimal maintenance planning and timetable adaption including

- an evaluation of the impacts on users' demand satisfaction;
- addressing the issue of service interruptions due to maintenance considering possible solutions outside the rail system environment. Compared to [15-20], this study deals with service interruptions with high duration and suggests alternative transport solutions thanks to the integration with other transport modes;
- dealing with the problem of service interruptions from the service provider point of view. For example, in comparison to [21], this means not just developing trip planning models capable of suggesting the best path to passengers, but designing alternative transport solutions that can mitigate the impact of service interruption both for users and operators;
- supporting the operator in evaluating the convenience of activating new services taking into account the activation costs and the transport demand that could remain unmet. Compared to [22-26], the current study models a cooperation among the different transport service providers extending the capabilities of bus bridging; compared to [27-29], it includes also the possibility of upgrading and integrating existing bus services;
- introducing the cooperation among different operators, such as metro and bus service operators, and a collaborative joint management of planned service interruptions, to minimize the impact on the multimodal journey. In more detail, compared to [30-33], it introduces a wider cooperative framework and a negotiation algorithm to mimic the real decisional process.

III. COOPERATIVE FRAMEWORK AND METHODOLOGY

The presented model, as mentioned, represents an innovative methodology that supports rail service providers in dealing with planned service interruptions, reducing the impact on rail attractiveness, and minimizing service providers' costs and unmet passenger transport demand. Without losing generality and for the sake of clearness, hereafter, the metro transport mode will be considered the system to be maintained, and the bus service provider will be considered a unique representative of the other transport services.

A cooperative framework is applied to evaluate the optimal allocation of maintenance interventions to maintenance windows, to define the best action for mitigating service disruptions, and to design alternative transport services, if needed.

The involved actors are:

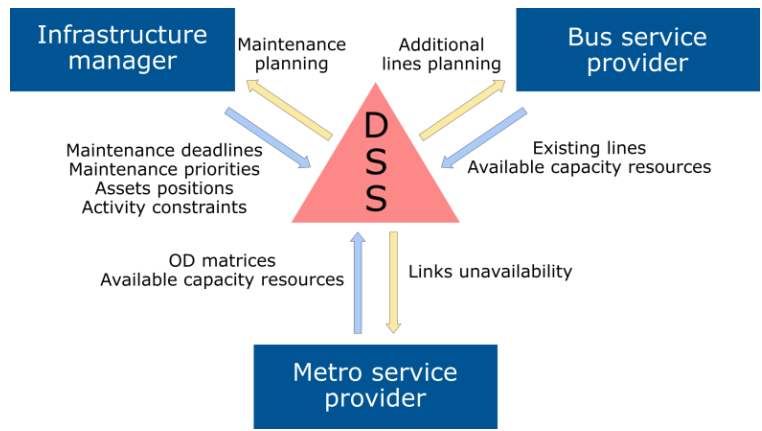

Fig. 1. The ideal cooperation scheme.

- the metro infrastructure manager/maintenance provider, who is in charge of planning maintenance interventions to guarantee infrastructure reliability;
- the metro service provider, whose goal is to guarantee a high level of service for passengers;
- the other service providers, which can accept some modifications to their service if this leads to some benefits, for example, in terms of number of passengers, but have to deal with limited resources.

These actors, which currently usually make decisions in an independent and isolated way, will be called on to cooperate to provide a multimodal service. The negotiation processes among the actors are quite complex in real-world scenarios, even if, as mentioned, in an urban environment the same company may be in charge of different transport modes, such as bus and metro services, making the negotiation process simpler.

An effective solution approach capable of taking into account these negotiations could be based on the scheme sketched in Fig. 1, where the decision support system embeds a mathematical programming model aimed at providing the following:

- the optimal allocation of metro infrastructure possessions to carry out maintenance interventions;
- the optimal evaluation of the needed actions, in terms of alternative node connections, to mitigate the impact of the interruption according to the transport demand;
- the design of the supplementary bus service and its integration with existing service.

To find a good shared solution, each actor provides a set of inputs and receives the relevant subset of the whole solution. In this framework, it is worth emphasizing that each solution subset is influenced by all three actors. In fact,

- Optimal maintenance planning is influenced by infrastructure manager activities (in particular, by the lateness of the maintenance activities with respect to their deadlines) and by the metro service provider (in particular, by the amount of unmet demand resulting from the rail line interruption);
- The metro service is influenced by the infrastructure manager (in particular, by the priorities and the deadlines of the different activities) and by the bus service provider (in particular, by its capability of providing, at a certain time, the required additional capacity);
- The problem of designing the supplementary bus connections provided by a third-party operator receives as input the requests of connections from the metro service provider and determines the optimal design of new lines. In doing so, the problem considers bus service provider's internal planning, with particular reference to its costs and residual capacity. Such a model could be included directly in the previous one whenever the metro service operator also manages the bus lines.

Nevertheless, the resulting problem turns out to be challenging to be exactly solved for real-world systems, especially if the metro and bus services must be adapted according to mobility demand dynamics. Therefore, since it requires the definition of an ad hoc solution algorithm, the cooperation scheme is decomposed, as in the scheme presented in Fig. 2, where the whole problem is divided into three subproblems, each associated with one of the actors.

Since the reciprocal influence of the three actors must be preserved, the three resulting subproblems are solved iteratively following the scheme depicted in Fig. 2, where each subproblem communicates the solution to the next subproblem (violet arrows), whereas subproblem 1 receives the effects of the other two subproblems (green arrows). In this framework, such a closed loop mechanism introduces a *negotiation process* among the three actors aimed at finding the best compromise between their different targets and key performance indicators, such as the deadline fulfillment, the users' unmet demand reduction, the capacity raising costs, and so on, which will be better described in the following sections.

In addition, this decomposed scheme allows to solve, for each solution of the first subproblem, the second and third subproblems for different times of the day (i.e., during the morning peak, evening peak, and off-peak of demand) of each week in the considered time horizon taking into account the demand dynamics. Therefore, it is possible to evaluate the effects of the same maintenance plan and infrastructure possession considering daily, weekly, and seasonal variations of passenger flows.

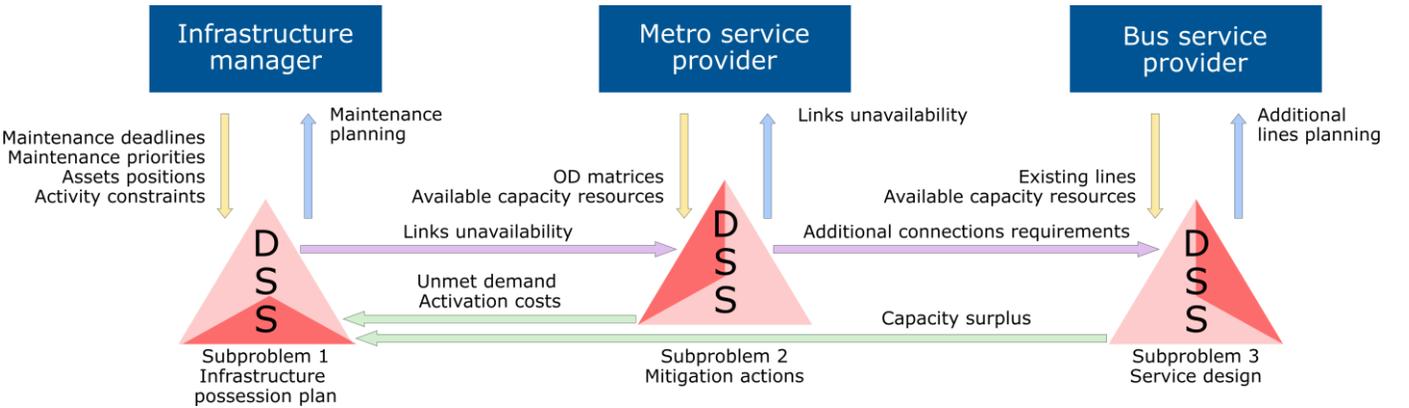

Fig. 2. The decomposition approach to the solution.

To conclude, it is possible to emphasize that the proposed decomposition allows the application of the three subproblems independently to deal with real-world situations in which part of the decisions have already been made. Significant examples consist of the cases when the maintenance plan is already defined, when new information regarding transport demand implies re-evaluating the second and third subproblems, and when a change in resource availability affects the bus service design.

IV. MATHEMATICAL FORMULATION

In this section, the mathematical formulation of the proposed model is presented. Specifically, the model is composed of three Mixed-Integer Linear Programming (MILP) problems according to the cooperative framework described above. In doing so, the three MILP subproblems are aimed at allocating maintenance interventions to specific weeks in the considered time horizon, at defining the best mitigation strategy, and at designing the alternative service.

The considered multimodal network is modeled as a graph $\mathcal{G} = \{\mathcal{N}, \mathcal{L}\}$, where $\mathcal{N}$ is the set of nodes and $\mathcal{L}$ is the set of links. It is assumed that $\mathcal{G}$ can be split into:

- the metro rail subgraph $\mathcal{G}_R = \{\mathcal{N}_R, \mathcal{L}_R\}$;
- the bus subgraph $\mathcal{G}_B = \{\mathcal{N}_B, \mathcal{L}_B\}$;
- the pedestrian subgraph $\mathcal{G}_W = \{\mathcal{N}_W, \mathcal{L}_W\}$;

6- the subgraph $\mathcal{G}_S = \{\mathcal{N}_S, \mathcal{L}_S\}$ of potentially activatable services.

By definition, such subgraphs are characterized by the following properties: $\mathcal{N} = \mathcal{N}_R \cup \mathcal{N}_B \cup \mathcal{N}_W \cup \mathcal{N}_S$, $\mathcal{L} = \mathcal{L}_R \cup \mathcal{L}_B \cup \mathcal{L}_W \cup \mathcal{L}_S$. However, while the resulting subsets of links are disjointed, a "multimodal node" can belong to different subsets of nodes (i.e., a node can have an input pedestrian link and an output bus link, belonging, therefore, to both $\mathcal{N}_W$ and $\mathcal{N}_B$). The description and the main assumptions of each subproblem are provided in the following subsections, where the superscript $r$ refers to the tabu search iteration described in Sec. V, where its meaning will be described in detail. Therefore, in the following subproblem description, it can always be considered a constant.

### A. First subproblem: Maintenance windows allocation

The first subproblem (SP1) is applied only to the metro rail subgraph $\mathcal{G}_R$. Given the set of metro rail links $\mathcal{L}_\mathcal{A} \subseteq \mathcal{L}_R$ needing maintenance, the optimal allocation of the interventions to time intervals is evaluated considering passenger transport demand and the operational constraints due to network topology and traffic management rules.

The assumptions of the first subproblem are as follows:
- the problem is solved in advance at a tactical level; therefore, the solution is determined offline;
- with the terms "maintenance intervention", a group of maintenance activities to be executed on the same metro rail link is considered as a whole since the goal of SP1 is only to determine the period of line possession during which the different activities are performed and train circulation is forbidden. The detailed schedule of maintenance activities within each week and their assignment to the available resources are outside the scope of this work;
- each maintenance intervention cannot be interrupted, and the link is closed until the end of the maintenance intervention;
- a discretization of the time horizon in weeks is considered. The time horizon is given by a set $\mathcal{T}$ of weeks, large enough to guarantee that all the interventions can be performed;
- all the interventions are planned at a time considering the whole set of weeks $\mathcal{T}$;
- the durations of the maintenance interventions are expressed in number of weeks, fixed and known in advance;
- at maximum $N$ links that can be interrupted at the same time.

Therefore, given the set of maintenance interventions $\mathcal{A}$ that have to be executed, each maintenance intervention $i \in \mathcal{A}$ is characterized by:
- a metro link $(n, m) \in \mathcal{L}_\mathcal{A} \subseteq \mathcal{L}_R$ where the intervention is located;
- a priority $\omega_i$, related to the criticality of the link for the rail network;
- a deadline $\theta_i$ that is the time limit within which the intervention must be finished;

TABLE I
Notation of subproblem 1.

| | |
|---|---|
| **SETS** | |
| $\mathcal{A}$ | Set of maintenance interventions (links needing maintenance) and $|\mathcal{A}|$ its cardinality |
| $\mathcal{N}_R$ | Set of nodes |
| $\mathcal{L}_R$ | Set of metro rail links |
| $\mathcal{G}_R = \{\mathcal{N}_R, \mathcal{L}_R\}$ | Graph representing the considered railway network |
| $\mathcal{L}_\mathcal{A}$ | Subset of $\mathcal{L}_R$ gathering the links with maintenance interventions |
| $\mathcal{T}$ | Set of weeks available for maintenance interventions; $|\mathcal{T}|$ identifies the last week in the considered period |
| **CONSTANTS** | |
| $\pi_i$ | Processing time of maintenance intervention $i$, expressed in number of weeks |
| $g^k$ | Network utilization rate in week $k$ |
| $\theta_i$ | Deadline for the maintenance intervention $i$ |
| $\Theta_{(n,m)}$ | Maximum deadline of the intervention to be executed along the link $(n,m) \in \mathcal{L}_R$ |
| $N$ | Maximum number of disrupted links at the same time |
| $\omega_i$ | Priority of maintenance intervention $i$ |
| $\alpha_j$ | Weight of the objective function $j$-th term; chosen according to service provider intentions |
| **MATRICES** | |
| $I$ | Inclusion matrix: its entry $I_{i,(n,m)}$ is equal to 1 if intervention $i$ has to be executed on link $(n, m) \in \mathcal{L}_R$ and to 0 otherwise; based on the following definition: $\mathcal{L}_\mathcal{A} = \{(n,m) \in \mathcal{L}_R : \sum_{i \in \mathcal{A}} I_{i,(n,m)} > 0\}$ |
| **VARIABLES** | |
| $t_i^r$ | Starting time of maintenance intervention $i$ at the $r$-th iteration |
| $x_{(n,m)}^{r,k}$ | Binary variable equal to 1 if the link $(n, m) \in \mathcal{L}_R$ is interrupted in the week $k$ of the $r$-th iteration and 0 otherwise |
| $y_i^{r,k}$ | Binary assignment variable equal to 1 if the week $k$ of the $r$-th iteration is the first for the intervention $i \in \mathcal{A}$ and 0 otherwise |
| $\boldsymbol{x}^r, \boldsymbol{y}^r$ | Vectors gathering the variables $x_{(n,m)}^{r,k}$ and $y_i^{r,k}$ |
| $\Lambda(\cdot)$ | Tabu term; linear combination of the variables in $\boldsymbol{x}^r$ and their values assumed in previous iterations |

- the number of weeks $\pi_i$ required to perform the maintenance intervention that needs to be performed on that link.

The network utilization rate $g^k$ in each week $k \in \mathcal{T}$ of the considered time horizon is known, according to the seasonal variation of the transport demand. Since each time interval is characterized by a given transport demand, a higher penalty is paid if a week with high transport demand is allocated to maintenance.

The decision to be made consists of the definition of the optimal allocation of maintenance interventions to time intervals and of the order in which the metro links need to be maintained. The objective function aims at minimizing the weighted sum of completion time $t_i^r$ of the maintenance interventions, evaluated also considering the maintenance intervention priority $\omega_i$, and of the network disruption, evaluated considering the utilization rate $g^k$.

The constraints ensure that all the needed maintenance interventions are executed before the deadline and within the considered time horizon to guarantee the proper infrastructure condition. Moreover, constraints force solutions in which adjacent segments must be closed at the same time to avoid the simultaneous interruption of too many links or of links located in different parts of the network.

Regarding the mathematical formulation, given the notation in Tab. I, the first subproblem is stated as

$$\min \alpha_1 \sum_{i \in \mathcal{A}} \omega_i (t_i^r + \pi_i) + \alpha_2 \sum_{k \in \mathcal{T}} \sum_{(n,m) \in \mathcal{L}_\mathcal{A}} g^k x_{(n,m)}^{r,k} + \alpha_3 \Lambda(\cdot) \qquad (1)$$

subject to:

$$t_i^r = \sum_{k \in \mathcal{T}} k y_i^{r,k} \qquad \forall i \in \mathcal{A} \qquad (2)$$

$$t_i^r + \pi_i \leq |\mathcal{T}| \qquad \forall i \in \mathcal{A} \qquad (3)$$

$$\sum_{k=\Theta_{(n,m)}+1} x_{(n,m)}^{r,k} = 0 \qquad \forall i \in \mathcal{A}, \forall (n,m) \in \mathcal{L}_\mathcal{A}, \Theta_{(n,m)} = \max_{\forall i: I_{i,(n,m)}=1} \theta_i \qquad (4)$$

$$\sum_{(n,m) \in \mathcal{L}_R} x_{(n,m)}^{r,k} \leq N \qquad \forall k \in \mathcal{T} \qquad (5)$$

$$x_{(n,m)}^{r,k} + x_{(l,h)}^{r,k} \leq 1 \qquad \forall k \in \mathcal{T}, \forall (n,m), (l,h) \in \mathcal{L}_\mathcal{A}: l \neq m \qquad (6)$$

$$x_{(n,m)}^{r,v} \leq 1 - y_i^{r,k} \qquad \begin{array}{l} \forall i \in \mathcal{A}, \forall (n,m) \in \mathcal{L}_\mathcal{A}: I_{i,(n,m)} = 1 \\ \forall v, k \in \mathcal{T}: v \leq k-1, k \in \mathcal{T}, \end{array} \qquad (7)$$

$$x_{(n,m)}^{r,v} \leq 1 - y_i^{r,k} \qquad \begin{array}{l} \forall i \in \mathcal{A}, \forall (n,m) \in \mathcal{L}_\mathcal{A}: I_{i,(n,m)} = 1 \\ \forall v, k \in \mathcal{T}: v \geq k + \pi_i, v \leq |\mathcal{T}|, \end{array} \qquad (8)$$

$$x_{(n,m)}^{r,v} \geq y_i^{r,k} \qquad \begin{array}{l} \forall i \in \mathcal{A}, \forall (n,m) \in \mathcal{L}_\mathcal{A}: I_{i,(n,m)} = 1 \\ \forall v, k \in \mathcal{T}: k-1 < v < k + \pi_i, \end{array} \qquad (9)$$

$$\sum_{k \in \mathcal{T}} y_i^{r,k} = 1 \qquad \forall i \in \mathcal{A} \qquad (10)$$

$$x_{(n,m)}^{r,k} = 0 \qquad \forall k \in \mathcal{T}, \forall (n,m) \in \mathcal{L}_R \setminus \mathcal{L}_\mathcal{A} \qquad (11)$$

where:
- the objective function in (1) formalizes the abovementioned SP1 goal and takes into account the tabu term $\Lambda(\cdot)$ representing the contribution provided by the second and third subproblems to quantify the effect of the solution of this subproblem on them. While a detailed discussion will be provided in Sec. V, it is sufficient to specify here that it is linear in the optimization variables $x^r$;
- the constraints in (2) define the initial time of each maintenance intervention;
- the constraints in (3) define that all the interventions must be completed in the considered time horizon;
- the constraints in (4) guarantee that the maintenance intervention $i$ is completed before its deadline;
- the constraints in (5) define the maximum number of links (maintenance interventions) that can be interrupted (executed) in the same week $k$;
- the constraints in (6) guarantee that two nonadjacent links cannot be interrupted in the same week $k$;
- the constraints in (7), (8) and (9) guarantee that the maintenance intervention on a link cannot be allocated to nonconsecutive time intervals (preemption not allowed). In detail:
    - the constraints in (7) guarantee that a link cannot be interrupted before the beginning of the maintenance intervention;
    - the constraints in (8) guarantee that a link cannot be interrupted after the final time interval;
    - the constraints in (9) guarantee that the link interruption lasts until the maintenance intervention is concluded (preemption not allowed);
- the constraints in (10) define that each intervention must have exactly one starting time interval in the considered time

horizon, meaning that all the maintenance interventions must be executed (or each link needing maintenance must be interrupted) in the considered time horizon;
- the constraints in (11) guarantee that the link cannot be interrupted if the considered maintenance interventions are not located in that link.

### B. Second subproblem: Mitigation strategy definition

In the second subproblem (SP2), the optimal mitigation strategy is determined separately for each period of day $b \in \mathcal{B}$ of each week $k$ in the considered time horizon $\mathcal{T}$ considering the passenger transport demand $d_{(o,d)}^{k,b}$ between each origin/destination pair $(o,d) \in \mathcal{P}$ and the service provider cost $c_{(n,m)}$ for activating a capacity unit $q_0$ on a new link $(n,m) \in \mathcal{L}_S$ or on an existing bus link $(n,m) \in \mathcal{L}_B$. To this aim, the entire multimodal network $\mathcal{G}$ is considered.

Regarding the input data, the list of links that will be interrupted in the considered time period is provided by the possession planning defined in SP1. In particular, $x_{(n,m)}^{r,k}$ is a constant equal to 1 if link $(n,m) \in \mathcal{L}_R$ is interrupted in week $k$ and 0 otherwise.

As regards SP2 hypotheses, it is assumed that:
- the passenger transport demand $d_{(o,d)}^{k,b}$ is known for all $(o,d) \in \mathcal{P}$ for all periods of day $b \in \mathcal{B}$ of each week $k \in \mathcal{T}$;
- The cost of a capacity activation $c_{(n,m)}$ is known $\forall (n,m) \in \mathcal{L}_S \cup \mathcal{L}_B$;
- The capacity of link $Q_{(n,m)}$ is known for each $(n,m) \in \mathcal{L} \setminus \mathcal{L}_S$;
- The generalized cost $\eta_{(n,m)}$, i.e., the combination of travel time and monetary cost, is known for each $(n,m) \in \mathcal{L}$;
- The maximum number of capacity units that can be activated, according to the service operator budget, is given;
- A deterministic network loading assignment is assumed to be representative since congestion and users' path choice randomness can be considered negligible because the metro and bus services are characterized by dedicated infrastructure or lanes and users are informed in advance regarding the best path, respectively.

Considering the cost function, the objective consists of minimizing the weighted sum of the generalized costs for passengers, in terms of travel time and pedestrian cost, and the cost for the service provider in terms of capacity activation cost and unmet demand.

Finally, the constraints guarantee the reachability and connection between each origin and destination by means of existing alternative services or ad hoc solutions and introduce a flow limitation according to the link capacity.

TABLE II
Notation of subproblem 2 (integrating Tab. I).

| | SETS |
|---|---|
| $\mathcal{N}$ | Set of nodes |
| $\mathcal{P}$ | Set of the pair origin-destination pairs: $\mathcal{P} = \{(o,d), o \in \mathcal{N}, d \in \mathcal{N}, o \neq d\}$ |
| $\mathcal{L}_B$ | Set of existing bus links whose capacity can be increased |
| $\mathcal{L}_R$ | Set of metro rail links |
| $\mathcal{L}_w$ | Set of pedestrian links |
| $\mathcal{L}_S$ | Set of activatable links |
| $\mathcal{L}$ | Complete set of links: $\mathcal{L} = \mathcal{L}_B \cup \mathcal{L}_R \cup \mathcal{L}_w \cup \mathcal{L}_S$ |
| $\mathcal{B}$ | Set of considered periods of the day |
| $\mathcal{D}$ | Set of o/d matrices for the different periods of the day in each week $k \in \mathcal{T}$ |
| $\mathcal{G} = \{\mathcal{N}, \mathcal{L}\}$ | Graph representing the considered multimodal network |
| $BS(n)$ | Backward Star of node $n \in \mathcal{N}$ |
| $FS(n)$ | Forward Star of node $n \in \mathcal{N}$ |
| | CONSTANTS |
| $r$ | Tabu search iteration counter |
| $D^{b,k}$ | Origin/Destination (o/d) matrix for the period $b \in \mathcal{B}$ of the day in the week $k \in \mathcal{T}, \forall b \in \mathcal{B}$ |
| $d_{(o,d)}^{k,b}$ | Entry of the matrix $D^{b,k}$ referring to the pair $(o,d) \in \mathcal{P}$ |
| $\eta_{(n,m)}$ | Generalized cost of the link $(n,m) \in \mathcal{L}$ |
| $q_{(n,m)}$ | Capacity of the existing link $(n,m) \in \mathcal{L} \setminus \mathcal{L}_S$ |
| $q_0$ | Capacity unit, that is, minimum number of seats that can be activated |
| $c_{(n,m)}$ | Cost for activating a new capacity unit on link $(n,l) \in \mathcal{L}_S \cup \mathcal{L}_B$ |
| $x_{(n,m)}^{r,k}$ | Binary constant equal to 1 if the link $(n,m) \in \mathcal{L}_R$ is interrupted in the week $k$ of the $r$-th iteration and 0 otherwise; these values are provided by the subproblem 1 |
| $\beta_j$ | Weight of the objective function $j$-th term, chosen according to metro service planner intentions |
| $M$ | Big M: integer suitably chosen to approximate $+\infty$ in the constraints |
| $J$ | Maximum number of capacity units to be activated |
| | VARIABLES |
| $h_{(n,l),(o,d)}^{r,k,b}$ | Path flow on the link $(n,m) \in \mathcal{L}$ for the pair $(o,d) \in \mathcal{P}$, in period $b$ of any day of the week $k$ |
| $f_{(n,m)}^{r,k,b}$ | Flow on the link $(n,m) \in \mathcal{L}$ in period $b \in \mathcal{B}$ of any day of the week $k \in \mathcal{T}$ |
| $w_{(n,m)}^{r,k,b}$ | Binary variable equal to 1 if the additional capacity is activated on the link $(n,m) \in \mathcal{L}_S$ in period $b \in \mathcal{B}$ of any day of week $k \in \mathcal{T}$, and 0 otherwise |
| $\varphi_{(o,d)}^{r,k,b}$ | Unmet demand for the pair $(o,d) \in \mathcal{P}$ in period $b \in \mathcal{B}$ of any day of week $k \in \mathcal{T}$ |
| $q_{(n,m)}^{r,k,b}$ | Capacity of the link $(n,m) \in \mathcal{L}$ in period $b \in \mathcal{B}$ of any day of week $k \in \mathcal{T}$ |
| $z_{(n,m),j}^{r,k,b}$ | Binary variable equal to 1 if the $j$-th capacity unit $q_0$ is activated on link $(n,m) \in \mathcal{L}_S \cup \mathcal{L}_B$ in period $b \in \mathcal{B}$ of any day of +week $k \in \mathcal{T}$, and 0 otherwise |

...

Concerning the mathematical formalization of the problem, it is worth remarking that subproblem 2 is formulated for a generic period of any day $b \in \mathcal{B}$ of a generic week $k \in \mathcal{T}$. Its solutions will be determined separately $\forall b \in \mathcal{B}$ and $\forall k \in \mathcal{T}$ via the algorithm reported in Sec. V. Therefore, considering the indices $b$ and $k$ as constants, the second subproblem is stated as

$$\min \beta_1 \sum_{(n,m)\in\mathcal{L}} \eta_{(n,m)} f^{r,k,b}_{(n,m)} + \beta_2 \sum_{(n,m)\in\mathcal{L}_S \cup \mathcal{L}_B} \sum_{j=1}^{J} c_{(n,m)} z^{r,k,b}_{(n,m),j} + \beta_3 \sum_{(n,m)\in\mathcal{L}_S} w^{r,k,b}_{(n,m)} + \beta_4 \sum_{(o,d)\in\mathcal{P}} \varphi^{r,k,b}_{(o,d)} \quad (12)$$

subject to:

$$\sum_{m\in FS(n)} h^{r,k,b}_{(n,m),(o,d)} - \sum_{m\in BS(n)} h^{r,k,b}_{(m,n),(o,d)} = d^{k,b}_{(o,d)} - \varphi^{r,k,b}_{(o,d)} \qquad \forall n \in \mathcal{N}, \forall (o,d) \in \mathcal{P}: o = n \quad (13)$$

$$\sum_{m\in FS(n)} h^{r,k,b}_{(n,m),(o,d)} - \sum_{m\in BS(n)} h^{r,k,b}_{(m,n),(o,d)} = -d^{k,b}_{(o,d)} + \varphi^{r,k,b}_{(o,d)} \qquad \forall n \in \mathcal{N}, \forall (o,d) \in \mathcal{P}: d = n \quad (14)$$

$$\sum_{m\in FS(n)} h^{r,k,b}_{(n,m),(o,d)} - \sum_{m\in BS(n)} h^{r,k,b}_{(m,n),(o,d)} = 0 \qquad \forall n \in \mathcal{N}, \forall (o,d) \in \mathcal{P}: o,d \neq n \quad (15)$$

$$f^{r,k,b}_{(n,m)} = \sum_{(o,d)\in\mathcal{P}} h^{r,k,b}_{(n,m),(o,d)} \qquad \forall (n,m) \in \mathcal{L} \quad (16)$$

$$f^{r,k,b}_{(n,m)} \leq q^{r,k,b}_{(n,m)} \qquad \forall (n,m) \in \mathcal{L} \quad (17)$$

$$q^{r,k,b}_{(n,m)} = q_0 \sum_{j=1}^{J} z^{r,k,b}_{(n,m),j} \leq M w^{r,k,b}_{(n,m)} \qquad \forall (n,m) \in \mathcal{L}_S \quad (18)$$

$$q^{r,k,b}_{(n,m)} = q_{(n,m)} + q_0 \sum_{j=1}^{J} z^{r,k,b}_{(n,m),j} \qquad \forall (n,m) \in \mathcal{L}_B \quad (19)$$

$$q^{r,k,b}_{(n,m)} = q_{(n,m)} \qquad \forall (n,m) \in \mathcal{L} \setminus (\mathcal{L}_B \cup \mathcal{L}_S) \quad (20)$$

$$q^{r,k,b}_{(m,n)} - q^{r,k,b}_{(n,m)} = 0 \qquad \forall (n,m) \in \mathcal{L}_S \quad (21)$$

$$f^{r,k,b}_{(n,m)} \geq 0 \qquad \forall (n,m) \in \mathcal{L} \quad (22)$$

$$f^{r,k,b}_{(n,m)} \leq M\bigl(1 - x^{r,k}_{(n,m)}\bigr) \qquad \forall (n,m) \in \mathcal{L}_\mathcal{A} \quad (23)$$

$$q^{r,k,b}_{(n,m)} \geq 0 \qquad \forall (n,m) \in \mathcal{L}_S \quad (24)$$

where:
- the objective function in (12) formalizes the above described SP2 goal;
- the constraints in (13), (14), and (15) represent the flow conservation equations for each node $n$ in case it coincides with the origin, with the destination, or in case it is none of the two, respectively;
- the constraints in (16) represent the link flow;
- the constraints in (17) represent the flow capacity constraints that guarantee that the capacity of the link cannot be exceeded;
- the constraints in (18), (19) and (20) define the capacity of each link. In particular,
    - the constraints in (18) define the capacity of the additional links if they are activated;
    - the constraints in (19) define the total capacity of the bus link, that is, the sum of the nominal, initial capacity and the additional activated units;
    - the constraints in (20) set the capacity of all the other links equal to the nominal one.
- the constraints in (21) ensure that the additional service is activated in both directions;
- the constraints in (22) guarantee that the flow cannot assume negative values;
- the constraints in (23) force a null value for the flow on the interrupted links;

- the constraints in (24) guarantee that the additional link capacity cannot assume negative values.

### C. Third subproblem: Service design

The third subproblem (SP3) aims to design, separately for each period of day $b \in \mathcal{B}$ of each week $k$ in the considered time horizon $\mathcal{T}$, the additional bus service considering only the subgraph of potential additional service $\mathcal{G}_S$.

Regarding SP3 input data, the activated additional links $(n,m) \in \mathcal{L}_S$ (i.e., those characterized by $w_{(n,m)}^{r,k,b} = 1$ in SP2 solution) and their optimal capacity $q_{(n,m)}^{r,k,b}$ are provided by SP2. It is worth nothing that the capacity $q_{(n,m)}^{r,k,b}$ is a requirement to meet, whereas the subproblem described here will provide the capacity $q_s^{r,k,b}$ of the activated/modified bus lines. In doing so, it is necessary that the capacity of line $s$ is equal to the maximum capacity of the links that belong to such a line. Nevertheless, since a link can belong to different lines, its required capacity $q_{(n,m)}^{r,k,b}$ can be provided by the sum of the capacities $q_s^{r,k,b}$ of the lines that include that link. In any case, the maximum number $S$ of services that can be activated is given according to the available budget and resources of the bus service provider.

In SP3, the decisions to be made are:
- the origin and the destination of each line $s$;
- the links that belong to each line $s$, identified by the binary variable $\omega_{(n,m),s}^{r,k,b} = 1$;
- the necessary capacity $q_s^{r,k,b}$ of each service $s$, expressed in terms of passengers/hour.

The objective is aimed at minimizing the service activation cost, properly dimensioning the bus lines by avoiding a surplus of line capacity with respect to the capacity that the bus line has to provide to the links it includes. To minimize the cost for the service operator, the objective function considers the minimization of the service capacity, the minimization of the total additional capacity provided by the activated services to the links and the minimization of the capacity surplus.

The considered constraints define the path of each activated line, guarantee a minimum capacity $q_0$ on each link of the service and define the number of bus lines that can be activated. Moreover, the constraints guarantee that the link capacity, defined by SP2 according to the transport demand, is satisfied.

TABLE III
Notation of subproblem 3 (integrating Tab. I and Tab. II).

| | SETS |
|---|---|
| $\mathcal{N}_S$ | Set of nodes |
| $\mathcal{L}_S$ | Set of activated additional links |
| $\mathcal{G}_S = \{\mathcal{N}_S, \mathcal{L}_S\}$ | Graph representing the additional network |
| $BS(n)$ | Backward Star of node $n \in \mathcal{N}_S$ |
| $FS(n)$ | Forward Star of node $n \in \mathcal{N}_S$ |
| | CONSTANTS |
| $q_{(n,m)}^{r,k,b}$ | Required capacity of the link $(n,m) \in \mathcal{L}_S$ provided by subproblem 2 in period $b$ of any day of week $k \in \mathcal{T}$ |
| $S$ | Number of activated services (bus lines) |
| $M$ | Integer suitably chosen to approximate $+\infty$ in the constraints |
| $v_j$ | Weight of the objective function $j$-th term, chosen according to service planner intentions |
| | VARIABLES |
| $\omega_{(n,m),s}^{r,k,b}$ | Binary variable equal to 1 if the link $(n,m) \in \mathcal{L}_S$ belong to service $s$ in period $b$ of any day of week $k \in \mathcal{T}$, and 0 otherwise |
| $q_s^{r,k,b}$ | Capacity of service $s$ during period $b$ of any day of week $k \in \mathcal{T}$ |
| $\mu_{(n,m),s}^{r,k,b}$ | Surplus of capacity; difference between $v_s$ and the capacity of the link $\varepsilon_{(n,m),s}^{r,k}$ provided by service $s$ in period $b$ of any day of week $k \in \mathcal{T}$ |
| $\varepsilon_{(n,m),s}^{r,k,b}$ | Capacity of the link $(n,m) \in \mathcal{L}_S$ provided by service $s$ in period $b$ of any day of week $k \in \mathcal{T}$ |
| $\xi_{n,s}^{r,k,b}$ | Binary variable equal to 1 if the node $n \in \mathcal{N}_S$ is the origin of service $s$ and 0 otherwise in period $b$ of any day of week $k \in \mathcal{T}$ |
| $\zeta_{n,s}^{r,k,b}$ | Binary variable equal to 1 if the node $n \in \mathcal{N}_S$ is the destination of service $s$ and 0 otherwise in period $b$ of any day of week $k \in \mathcal{T}$ |

It is worth noting that the detailed synchronization of the existing services and the detailed planning of the alternative services are outside the scope of the paper; therefore, the evaluation of transfer times to guarantee the connections is neglected; possible solutions for this fine tuning are available in the relevant literature [29], [30].

Analogously to SP2, SP3 is formulated for a generic period of any day $b \in \mathcal{B}$ of a generic week $k \in \mathcal{T}$. Its solutions will be determined separately $\forall b \in \mathcal{B}$ and $k \in \mathcal{T}$ via the algorithm reported in Sec. V. Regarding the mathematical formulation, given the notation in Tab. III and considering the indices $b$ and $k$ as constants, the second subproblem is stated as

$$\min v_1 \sum_{s=1}^{S} q_s + v_2 \sum_{s=1}^{S} \sum_{(n,m) \in \mathcal{L}_S} \varepsilon_{(n,m),s}^{r,k,b} + v_3 \sum_{s=1}^{S} \sum_{(n,m) \in \mathcal{L}_S} \mu_{(n,m),s}^{r,k,b} \quad (25)$$

subject to:

$$\sum_{m \in FS(n)} \omega_{(n,m),s}^{r,k,b} - \sum_{m \in BS(n)} \omega_{(n,m),s}^{r,k,b} = \xi_{n,s}^{r,k,b} - \zeta_{n,s}^{r,k,b} \qquad \forall n \in \mathcal{N}_S, \forall s = 1,..,S \qquad (26)$$

$$\xi_{n,s}^{r,k,b} + \zeta_{n,s}^{r,k,b} \leq 1 \qquad \forall n \in \mathcal{N}_S, \forall s = 1,..,S \qquad (27)$$

$$\sum_{s=1}^{S} \xi_{n,s}^{r,k,b} = 1 \qquad \forall n \in \mathcal{N}_S \qquad (28)$$

$$\sum_{s=1}^{S} \zeta_{n,s}^{r,k,b} = 1 \qquad \forall n \in \mathcal{N}_S \qquad (29)$$

$$\sum_{s=1}^{S} \varepsilon_{(n,m),s}^{r,k,b} \geq q_{(n,m)}^{r,k,b} \qquad \forall (n,m) \in \mathcal{L}_S \qquad (30)$$

$$q_s^{r,k,b} \geq max(\varepsilon_{(n,m),s}^{r,k,b}) \qquad \forall s = 1,..,S \qquad (31)$$

$$\varepsilon_{(n,m),s}^{r,k,b} \leq M \omega_{(n,m),s}^{r,k,b} \qquad \forall (n,m) \in \mathcal{L}_S, \forall s = 1,..,S \qquad (32)$$

$$\varepsilon_{(n,m),s}^{r,k,b} \leq q_0 - M(1 - \omega_{(n,m),s}^{r,k,b}) \qquad \forall (n,m) \in \mathcal{L}_S, \forall s = 1,..,S \qquad (33)$$

$$\mu_{(n,m),s}^{r,k,b} = q_s^{r,k,b} - \varepsilon_{(n,m),s}^{r,k,b} \qquad \forall (n,m) \in \mathcal{L}_S, \forall s = 1,..,S \qquad (34)$$

$$\omega_{(n,m),s}^{r,k,b} - \omega_{(m,n),s}^{r,k,b} = 0 \qquad \forall n,m \in \mathcal{N}_S, \forall s = 1,..,S \qquad (35)$$

where:
- the objective function in (25) formalizes the above described SP3 goal;
- the constraints in (26-29) define the path covered by each line considering that the origin and the destination are optimally chosen by the model; in detail:
  - the constraints in (26) define the path from the origin to the destination of the line $s$. In doing so, if $n$ is the origin of the line $s$ then $\xi_{n,s}^{k,b} = 1$ and if $n$ is the destination of the line $s$ then $\zeta_{n,s}^{k,b} = 1$;
  - the constraints in (27) define that the same node cannot be the origin and destination of the line;
  - the constraints in (28) and (29) define that each line has only one origin and one destination;
- the constraints in (30) guarantee that the sum of the capacity provided by the lines gathering a generic link $(n,m)$ is at least equal to the capacity required by SP2 for such a link;
- the constraints in (31) define the capacity of the line $s$, as the maximum capacity of the link that belongs to that line itself;
- the constraints in (32) guarantee that if link $(n,m)$ does not belong to service $s$, the capacity activated in that link for that service is null;
- the constraints in (33) state that if link $(n,m)$ belongs to line $s$, then the capacity activated in that link should be at least equal to the minimum service capacity $q_0$;
- the constraints in (34) define the surplus of service capacity with respect to capacity $\varepsilon_{(n,m),s}^{r,k,b}$ that the line $s$ has to provide to the link;
- the constraints in (35) ensure that the additional service is activated in both directions.

V. THE SOLUTION APPROACH

As mentioned, the solution approach is based on the problem decomposition and on a negotiation algorithm, which follows the cooperative approach described in Fig. 2 thanks to a "weighted tabu search" hereafter described. To suit to the proposed solution approach, the above introduced subproblems have been designed to exchange the necessary inputs and outputs with each other, as pointed out in Fig. 2.

*A. Definition of the tabu search term*

The negotiation mechanism is based on a "weighted tabu search", which consists of considering in the objective function of the first subproblem the performances of the other subproblems via the term $\Lambda(r, \boldsymbol{x}^r, \boldsymbol{\varphi}^{r-1}, \boldsymbol{c}^{r-1}, \boldsymbol{\mu}^{r-1}, \boldsymbol{X}^{r-1}, \mathcal{D})$, which is a function of the iteration $r$ and of:
- the present optimization variables $\boldsymbol{x}^r$, that is, the variables of subproblem 1 in the $r$-th iteration;
- the set of total unmet demand $\boldsymbol{\varphi}^{r-1} = \{\varphi^p : \forall p = 1, ... r - 1\}$ previously determined by SP2;
- the set of activation cost $\boldsymbol{c}^{r-1} = \{c^p : \forall p = 1, ... r - 1\}$ previously determined by SP2;

- the set of capacity surplus $\boldsymbol{\mu}^{r-1} = \{\mu^p: \forall p = 1, \ldots r-1\}$ previously determined by SP3;
- the set of solutions $\boldsymbol{\mathcal{X}}^{r-1} = \{\boldsymbol{x}^p: \forall p = 1, \ldots r-1\}$ previously determined by SP1;
- the set $\mathcal{D} = \{D^{b,k}: \forall b \in \mathcal{B}, \forall k \in \mathcal{T}\}$ of all the o/d matrices.

The function $\Lambda(\cdot)$ is chosen to act as a repulsion term that helps to explore new solutions with respect to the previously determined solutions. For what concerns its definition, it has been chosen as

$$\Lambda(r, \boldsymbol{x}^r, \boldsymbol{\varphi}^{r-1}, \boldsymbol{c}^{r-1}, \boldsymbol{\mu}^{r-1}, \boldsymbol{\mathcal{X}}^{r-1}, \mathcal{D}) = \sum_{k \in \mathcal{T}} \sum_{p=0}^{r-1} \gamma^p \sum_{\forall (n,m) \in \mathcal{L}_S} \delta\big(x_{(n,m)}^{p,k}, x_{(n,m)}^{r,k}\big) \tag{36}$$

where $x_{(n,m)}^{p,k}$ is the solution provided for link $(n,m) \in \mathcal{L}_R$ by SP1 in a generic iteration $p$, $\delta\big(x_{(n,m)}^{p,k}, x_{(n,m)}^{r,k}\big)$ is the Kronecker impulse and $\gamma^p$ is a weight chosen according to the model described below. With this choice, if $x_{(n,m)}^{p,k} = x_{(n,m)}^{r,k}$, that is, if the actual solution $x_{(n,m)}^r$ is equal to the $p$-th previous solution, there is a positive contribution to $\Lambda(\cdot)$; otherwise, the contribution is null. In addition, depending on the value of $\gamma^p$, the contributions given by $\delta\big(x_{(n,m)}^{p,k}, x_{(n,m)}^{r,k}\big)$, $\forall (n,m) \in \mathcal{L}_R$, can be amplified or reduced. To this aim, let

$$\gamma^p = \frac{\chi^p + c^p + \mu^p}{r - p}, \qquad \forall p = 1, \ldots, r-1 \tag{37}$$

where:

- $\chi^p = \sum_{\forall k \in \mathcal{T}} \sum_{\forall b \in \mathcal{B}} \sum_{\forall (o,d) \in \mathcal{P}} \varphi_{(o,d)}^{p,k,b} / d_{(o,d)}^{k,d}$ is the percentage of unmet demand with respect to the total demand determined by the solution of iteration $p$;
- $c^p = \sum_{j=1}^{J} \sum_{\forall (n,m) \in \mathcal{L}_S} \sum_{\forall b \in \mathcal{B}} \sum_{\forall k \in \mathcal{T}} c_{(n,m)} z_{(n,m),j}^{p,k,b}$ is the total cost of service activation determined by the solution of iteration $p$;
- $\mu^p = \sum_{\forall k \in \mathcal{T}} \sum_{b \in \mathcal{B}} \sum_{\forall (n,m) \in \mathcal{L}_S} \mu_{(n,m)}^{p,k,b}$ is the total capacity surplus determined by the solution of iteration $p$.

With this choice, high values of the above terms determine a high weight $\gamma^p$, increasing the weight of $\delta\big(x_{(n,m)}^{p,k}, x_{(n,m)}^{r,k}\big)$ in $\Lambda(\cdot)$, and then in Eq. (1). In addition, recent solutions ($p$ near $r$) weigh more than older solutions.

TABLE IV
Negotiation Algorithm.

| | |
|---|---|
| **Initialization** | Set $r = 1$ and let $R$ be the maximum number of refinement iterations |
| | Set $x_{(n,m)}^{k,0} = 0, \forall (n,m) \in \mathcal{L}_R, \chi^0 = 0, c^0 = 0, \mu^0 = 0$ |
| | **while** $r \leq R$ |
| **Step 1.** | Determine the tabu term $\Lambda(r, \boldsymbol{x}^r, \boldsymbol{\varphi}^{r-1}, \boldsymbol{c}^{r-1}, \boldsymbol{\mu}^{r-1}, \boldsymbol{\mathcal{X}}^{r-1}, \mathcal{D})$ |
| | **Solve the SP1** |
| **Step 2.** | For $k = 1, \ldots, |\mathcal{T}|$ |
| |    for $b = 1, \ldots, |\mathcal{B}|$ |
| |      By considering the solution of the $r$-th instance of SP1, **solve SP2** for period $b$ of the day and for week $k$ |
| |      By considering the solution of the $r$-th instance of SP1 and SP2, **solve SP3** for period $b$ of the day and for week $k$ |
| |    end for |
| | end for |
| **Step 3.** | Compute $\chi^r$, $c^r$, $\mu^r$, and $\gamma^r$ |
| | Set $r = r + 1$ |
| | **end while** |
| **Stop.** | Choose the best solution among those found during the $R$ iterations |

To conclude, it is worth noting that since $x_{(n,m)}^{p,k}$ and $x_{(n,m)}^{r,k}$ are binary, then $\delta\big(x_{(n,m)}^{p,k}, x_{(n,m)}^{r,k}\big) = 1 - \big|x_{(n,m)}^{p,k} - x_{(n,m)}^{r,k}\big|$ and the term in (36) are linear with respect to the variables $x_{(n,m)}^{r,k}$.

*B. Solution algorithm*

Regarding the solution algorithm, the three subproblems are solved iteratively, and at each iteration, the tabu term $\Lambda(\cdot)$ is updated according to its definition described in Sec. V.A.

The complete proposed solution algorithm is reported in Tab. IV, where it is worth noting that SP1 is solved once for all the weeks in $\mathcal{T}$, whereas SP2 and SP3 are solved separately for each week $k \in \mathcal{T}$ and for each period of the day $b \in \mathcal{B}$.

In more detail, after the initialization, the solution approach consists of iteration over a sequence of instances of the aforementioned three subproblems suitably stated. In particular:

- in Step 1, the tabu term is determined and the first subproblem is solved, evaluating the optimal allocation of maintenance interventions to time intervals for the iteration $r$.
- in Step 2, the second subproblem is solved and the additional links to be activated are identified as well as their capacity.

Moreover, by considering the solution of the second subproblem, the third subproblem is solved and the activated services are designed, in terms of links assignment to each bus line and definition of service capacity. Such a step is iterated over the set of considered transport demand matrices $\mathcal{B}$ and over the set of weeks $\mathcal{T}$.

- in Step 3, the percentage of unmet demand, the total cost of service activation and the total capacity surplus are computed for the iteration $r$.

All the steps are iterated until a maximum number $R$ of iterations is reached. After the last iteration, the solution identified as the best one in SP2 is assumed as the one to be applied. Such a choice is determined by the fact that SP2 is the most significant subproblem from the point of view of user satisfaction.

Concerning the computational effort, the solution time can be expressed as $R(AST_1 + |\mathcal{B}| \cdot |\mathcal{T}| \cdot (AST_2 + AST_3))$, being $AST_i$ the Average Solution Time (AST) of the generic subproblem $i$. Nevertheless, since SP2 is solved many times and, in general, is the most time-consuming, it turns out that the total computational effort is $\simeq R \cdot |\mathcal{B}| \cdot |\mathcal{T}| \cdot AST_2$, as confirmed by the case study analyses.

## VI. CASE STUDIES

In this section, two real-world case studies are implemented and analyzed with the aim of evaluating the performance of the proposed model.

The three subproblems (Sec. IV) and the solution algorithm (Sec. V) were implemented via IBM-Ilog Cplex® and Matlab®, respectively, on a 3.10 GHz PC with 16 GB RAM.

Each case study was tested with two different tabu search settings corresponding to a maximum number of refinement iterations $R = 10$ and $R = 20$. Finally, all three subproblems have been solved exactly, and all the performed analyses have been averaged considering 15 problem instances, each characterized by randomly generated o/d matrices. In this connection, the entries of the o/d matrices were sampled from a Gaussian distribution with the expectation equal to nominal $d_{(o,d)}^{k,b}$ and standard deviation equal to 10% of $d_{(o,d)}^{k,b}$. The weight coefficients of the cost functions, chosen to make the terms comparable, are the same for the two case studies and are reported in Tab. V.

### A. First case study: Genoa, Italy

The first case study is focused on the multimodal urban transport network of the Italian city of Genoa. The considered multimodal network consists of a metro line to be maintained, of the urban stretch of the regional/national rail line, and of four bus lines. The map of the considered portion of the city is depicted in Fig. 3.

The considered network is characterized by a graph with 21 nodes and 122 links where, thanks to the existing unified ticketing policy, the bus service and the urban section of ground rail service act together as a unique alternative transport mode with respect to the metro line. Detailed information regarding the graph size is reported in Tab. VI, while the graph is depicted in Fig. 4.

A time period of 20 weeks is considered. Nine metro links need maintenance (links to be interrupted) in the considered time period, and the durations of the necessary interventions of one or two weeks are known. The priorities and deadlines are defined for each maintenance intervention. In particular, the priorities are related to the criticality of the interventions in terms of the impact on service reliability and are expressed as a weight from 1 to 10. The deadlines of the interventions are expressed in terms of the remaining weeks before the line section reaches an unacceptable condition. The information regarding the needed maintenance interventions is reported in Tab. VII.

The number of available resources for performing maintenance is given, and the maximum number of links that can be interrupted at the same time is equal to three links. The bus timetable is known in advance and is assumed to be unaffected by disruptions.

Regarding the transport demand dynamics, two main features are modeled, as shown in Fig. 5:

- a forecasted peak during the 14th week due to an exceptional event (e.g., international exhibition);
- a seasonal variation occurs between weeks 10 and 11 (for instance triggered by the end of schools and the beginning of a holiday period).

According to the seasonal variation, weight $g_k$ is assigned to each week, which considers the forecast network utilization rate in that interval. The first 10 weeks present a network utilization rate twice the value of the network utilization rate of the last 10 weeks, with the exception of the 14th week during which there is the highest network utilization rate due to the abovementioned transport demand peak.

Regarding the daily variation, three o/d matrices are considered and are assumed to be known: the morning peak, the evening peak, and the off-peak transport demand. Their values are reported in the Appendix. The total demand of the considered portion of the Genoa public transport network is 17,100 passengers/hour during the morning peak, 21,600 passengers/hour during the evening peak and 5,800 passengers/hour during the off-peak. These values correspond to 25.5%, 32%, and 8.7% of the total transport peak demand of the city of Genoa, according to the official public transport o/d matrix provided by the Municipality of Genoa.

TABLE V
Cost function weight coefficients.

| Problem | Values | | | |
|---|---|---|---|---|
| SP1 | $\alpha_1 = 1$ | $\alpha_2 = 50$ | $\alpha_3 = 100$ | |
| SP2 | $\beta_1 = 0.03$ | $\beta_2 = 1.5$ | $\beta_3 = 1$ | $\beta_4 = 1$ |
| SP3 | $v_1 = 1$ | $v_2 = 1$ | $v_3 = 1$ | |

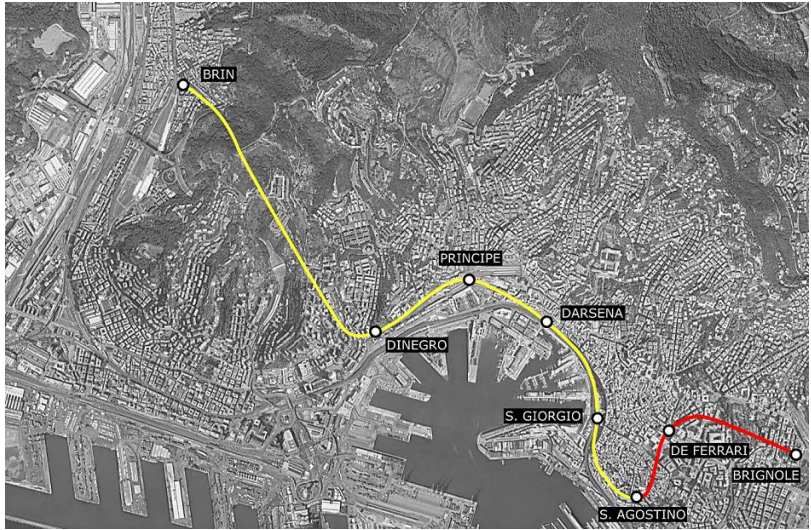

Fig. 3. The considered Genoa metro line (color legend: yellow – single direction to be interrupted for maintenance; red – both directions to be interrupted for maintenance).

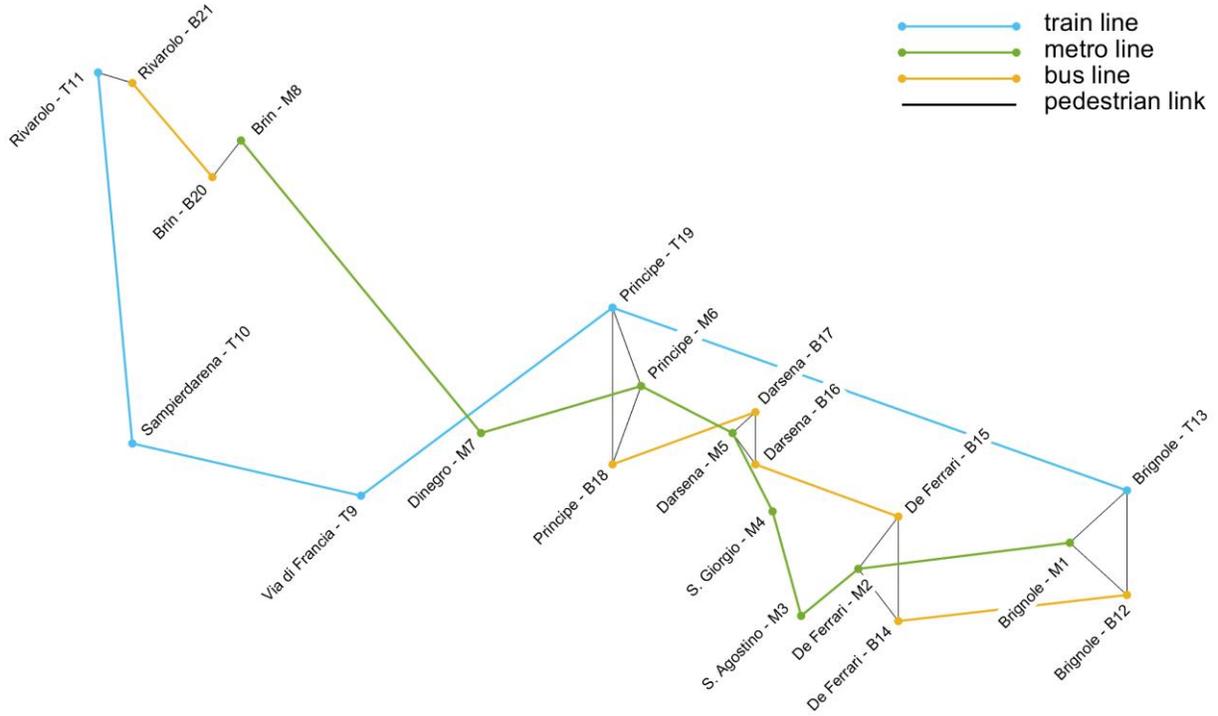

Fig. 4. The existing Genoa multimodal network (without the activatable links).

TABLE VI
Genoa graph sizes.

| NODES | |
|---|---|
| Number of metro nodes | 8 |
| Number of alternative service nodes | 13 |
| *Total number of nodes* | *21* |
| LINKS | |
| Number of metro links | 14 |
| Number of alternative service links | 16 |
| Number of pedestrian links | 28 |
| Number of additional links | 64 |
| *Total number of links* | *122* |

TABLE VII
Maintenance interventions.

| Link ID | Section | Duration (weeks) | Priority | Deadline (weeks) |
|---|---|---|---|---|
| **1** | Brignole – De Ferrari | 1 | 2 | 20 |
| **2** | De Ferrari – Brignole | 2 | 1 | 20 |
| **3** | De Ferrari – S. Agostino | 1 | 10 | 10 |
| **4** | S. Agostino – De Ferrari | 2 | 4 | 20 |
| **5** | S. Agostino – San Giorgio | 1 | 8 | 10 |
| **7** | San Giorgio – Darsena | 2 | 7 | 10 |
| **9** | Darsena – Principe | 1 | 3 | 20 |
| **12** | Dinegro – Principe | 2 | 5 | 20 |
| **13** | Dinegro – Brin | 1 | 1 | 20 |

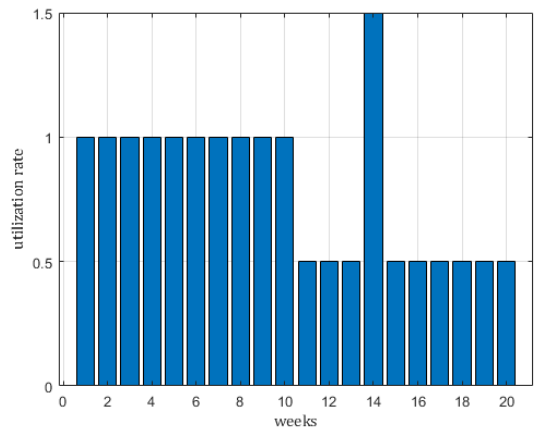

Fig. 5. Utilization rate (multiplier of the nominal demand value)

The sizes of the subproblem instances are reported in Tab. VIII, where it is worth noting that SP2 is significantly more complex than the other two subproblems since it considers the entire multimodal graph and the flow assignment to the network. With these parameters, the total resolution time turned out to be approximately 27 minutes with $R = 10$ and 53 minutes for $R = 20$. These solution times are definitively acceptable since the model is an offline tool used to solve tactical problems weeks in advance with respect to the planned service interruptions.

To evaluate the quality of the solutions, the obtained results have been compared with those achievable with two common maintenance strategies:
- the "position-based strategy", which consists of executing the interventions according to the link position (from link 1 to link 14).
- the "priority-based strategy", which implies the execution of maintenance interventions according to priorities. In the event of two or more interventions presenting the same priority, the sorting process considers the nearest as the next one.

TABLE VIII
Computational time averaged over 15 randomly generated problem instances.

|  | Average Solution Time (AST) [s] | Standard deviation [s] | Problem variables | Constraints |
|---|---|---|---|---|
| **SP1 (single instance)** | 0.25 | 0.03 | 482 | 6802 |
| **SP2 (single instance)** | 1.84 | 0.93 | 52260 | 61231 |
| **SP3 (single instance)** | 0.79 | 0.80 | 17737 | 18252 |
| **Overall problem – $R = 10$ iterations (each iteration implies solving SP2 and SP3 for 20 weeks and 3 o/d matrices)** | $1.61 \cdot 10^3$ | 0.36 | - | - |
| **Overall problem – $R = 20$ iterations (each iteration implies solving SP2 and SP3 for 20 weeks and 3 o/d matrices)** | $3.16 \cdot 10^3$ | 0.41 | - | - |

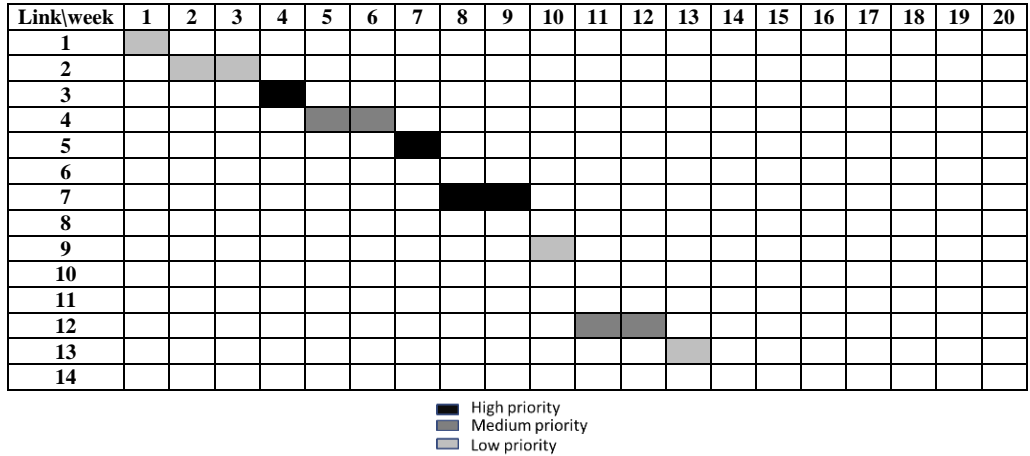

Fig. 6. Possession plan – planning strategy according to link position.

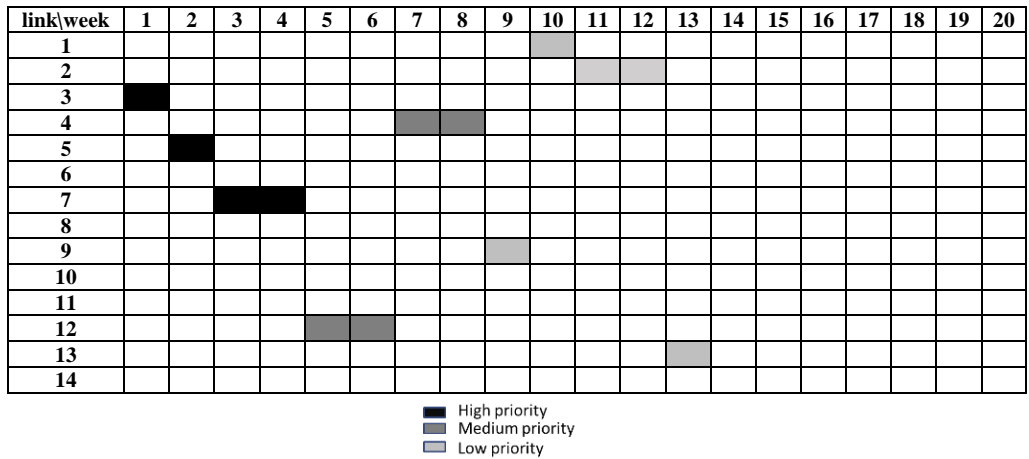

Fig. 7. Possession plan – planning strategy according to link priority.

In Fig. 6 and Fig. 7, the maintenance possession plans according to the two reference strategies are depicted for all 20 weeks considered, also indicating the intervention priorities. In both cases, the interventions are completed within week 13, but the transport demand and the impact on service are not considered.

Regarding the solution provided by SP1, the plans reported in Fig. 8 and Fig. 9 show the first solution (iteration $r = 1$) and the best solution found within $R$ iterations, respectively. In the first solution, there is no link interruption during weeks 6 to 10, 12, 14, and 20. A simultaneous interruption of links 3 and 4 is planned during week 3, since, in the same time interval, different links may be interrupted if they are adjacent. All interventions are scheduled before the relevant deadline, as guaranteed by SP1. Fig. 9 shows that in the best solution, there is no link interruption during weeks 5 to 10, 14 and 20. In particular, for week 14, this is due to the expected peak of demand. Finally, it is worth noting that links 1 and 2 are simultaneously maintained at week 17; these interventions have the same priority, whereas in the first solution, links 3 and 4 are simultaneously maintained although they had different priorities. Such different behaviors are allowed by the constraints.

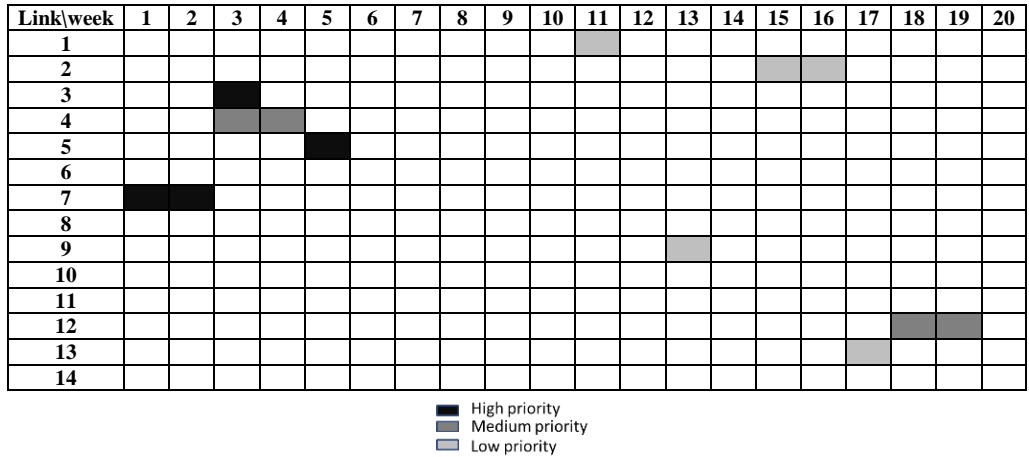

Fig. 8. Possession plan – first feasible solution ($r = 1$).

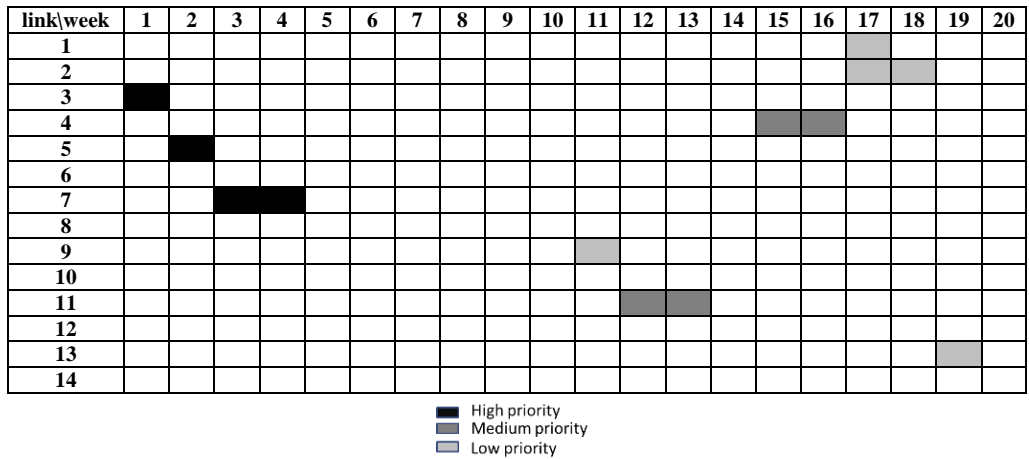

Fig. 9. Possession plan – best solution.

Regarding the solution performances, Tab. IX shows the difference between the best solution with respect to the first solution ($r = 1$) and the two reference strategies computed.

In particular, Tab. IX shows the main results for the infrastructure manager and for the service provider in terms of weighted completion time of interventions and impact on passengers according to the transport demand, respectively. In such a table, it is possible to note that the best solution guarantees significant improvements with respect to the position-based and priority-based strategies but also a good improvement with respect to the first solution ($r = 1$). As expected, compared to the priority-based strategy, the reduction of the impact on passengers in the best solution implies an increase in the weighted completion time. This increase does not occur in comparison with the position-based strategy in which the priority order is neglected and the weighted completion time is already characterized by a high value. In any case, a significant reduction of the impact on passengers is achieved in the best solution in comparison to both strategies.

TABLE IX
Solution comparison with respect to the best solution of SP1 (averaged over 15 randomly generated problem instances and $R = 10$).

| Objectives | Variation with respect to the first feasible solution | Variation with respect to the position-based strategy | Variation with respect to the priority-based strategy |
|---|---|---|---|
| Weighted maintenance completion time based on priority | -1% | -1% | +58.0% |
| Impact on passengers | -11% | -26% | -26% |
| Total | -8% | -21% | -15% |

TABLE X
Solution comparison with respect to the best solution of SP2 (averaged over 15 randomly generated problem instances and $R = 10$).

| | Objectives | Variation with respect to the first feasible solution | Variation with respect to the position-based strategy | Variation with respect to the priority-based strategy |
|---|---|---|---|---|
| **Morning Peak** | Alternative service activation cost | -72.4% | -80.0% | -85.7% |
| | Pedestrian links generalized cost | 3.0% | 2.2% | 3.4% |
| | Metro links generalized cost | 1.0% | 0.1% | 1.0% |
| | Bus links generalized cost | 2.7% | 2.4% | 0.1% |
| | Train links generalized cost | 1.0% | 1.5% | 1.7% |
| | Additional links generalized cost | -78.9% | -73.1% | -86.1% |
| | Unsatisfied demand cost | -29.9% | -81.3% | -68.2% |
| **Evening Peak** | Alternative service activation cost | -53.4% | -66.5% | -69.4% |
| | Pedestrian links generalized cost | 2.7% | 1.4% | 2.0% |
| | Metro links generalized cost | 5.0% | 4.8% | 2.4% |
| | Bus links generalized cost | 4.5% | 4.3% | 5.0% |
| | Train links generalized cost | 0.5% | 3.6% | 3.5% |
| | Additional links generalized cost | -51.4% | -46.1% | -48.3% |
| | Unsatisfied demand cost | -29.7% | -99.3% | -98.7% |
| **Off Peak** | Alternative service activation cost | -25.1% | -22.2% | -25.6% |
| | Pedestrian links generalized cost | -7.2% | -10.3% | -12.6% |
| | Metro links generalized cost | 3.0% | 5.6% | 7.8% |
| | Bus links generalized cost | -10.0% | -11.6% | -9.1% |
| | Train links generalized cost | -3.1% | -6.1% | -7.3% |
| | Additional links generalized cost | -25.1% | -23.6% | -26.1% |
| | Unsatisfied demand cost | 0.0% | 0.0% | 0.0% |

Tab. X shows, for SP2, the detailed differences of the best solution with respect to the first solution and to the two reference strategies, here considered as alternatives to the solution of SP1. In such a table, it is possible to note a significant reduction of the alternative service activation cost, of the unmet demand cost, and of the generalized cost of the additional activated links. On the other hand, the increase in the generalized cost on the pedestrian, metro, and alternative service links is negligible and is caused by the fact that, in the best solution, more users use alternative services. Regarding off-peak demand, the average variation in the unsatisfied demand cost is always null since the total demand is low and always totally satisfied.

Tab. XI lists the variation in the objective function terms of SP3 and, in particular, the needed average additional capacity to be activated for the new and existing bus lines and the average capacity surplus. In such a table, it is worth noting that the best solution provides significant improvements in comparison with all the reference solutions.

Details about the best solutions of SP2 and SP3 are provided in the Appendix, where sketches of the final supplementary links and lines to be activated are provided.

TABLE XI
Solution comparison with respect to the best solution of SP3 (averaged over 15 randomly generated problem instances and $R = 10$).

|  | Objectives | Variation with respect to the first feasible solution | Variation with respect to the position-based strategy | Variation with respect to the priority-based strategy |
|---|---|---|---|---|
| **Morning Peak** | Needed line capacity | -33.3% | -31.0% | -41.2% |
|  | Needed links capacity | -38.5% | -48.9% | -71.4% |
|  | Capacity surplus | -25.7% | -30.6% | -40.0% |
| **Evening Peak** | Needed line capacity | -13.8% | -10.6% | -20.0% |
|  | Needed links capacity | -45.3% | -32.1% | -17.4% |
|  | Capacity surplus | -11.1% | -9.1% | -20.1% |
| **Off Peak** | Needed line capacity | -29.3% | -18.0% | -33.2% |
|  | Needed links capacity | -47.7% | -71.3% | -72.6% |
|  | Capacity surplus | -28.9% | -15.2% | -31.4% |

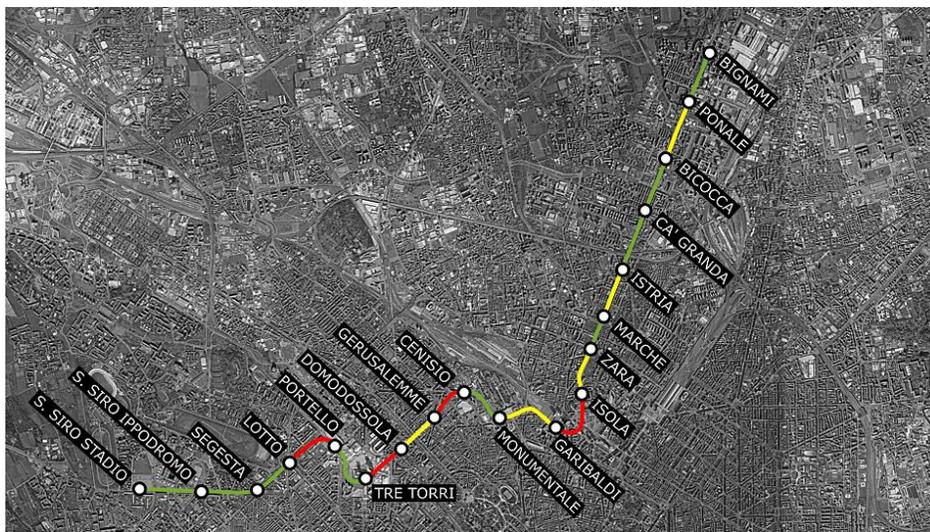

Fig. 10. The considered Milan metro line (color legend: green – no maintenance to plan; yellow – single direction to be interrupted for maintenance; red – both directions to be interrupted for maintenance).

In summary, on the one hand, the good performances achieved by the three subproblems confirm that the proposed solution algorithm is capable of balancing the performances by finding better final values for most of the objective functions. On the other hand, the computational effort required for the solution of the first case study is limited.

*B. Second case study: Milan, Italy*

To better assess the quality of the solution and computational performances, the proposed model and solution algorithms have also been tested on a larger network (Fig. 10) by considering a portion of the public transport network of the Italian city of Milan whose data are reported in Tab. XII and whose shape is depicted in Fig. 11.

The maintenance interventions are reported in Tab. XIII, whereas the o/d matrices are not reported due to space limitations.

Regarding the computational effort, the average solution times are reported in Tab. XIV, where it is possible to note that the increase in nodes and links (approximately 2 and 1.5 times the relevant values of the first case study, respectively) leads to a solution time of 3.65 hours with $R = 10$ and 7.50 hours for $R = 20$. In addition, also in this case, SP2 is confirmed as the most challenging to solve. In any case, these times remain acceptable since the proposed approach is applied offline, weeks in advance with respect to the planned service interruptions.

The different terms of the objective functions of the three subproblems are reported in Tabs. XV, XVI and XVII. The average values of SP1 are reported in Tab. XV ($R = 10$) by comparing the best solution with respect to the first solution ($r = 1$) and the two reference strategies.

Also in this case, such results demonstrate the good performance of the model in dealing with a larger instance of the problem, despite an increase in computational time. Analogous to the first case study, the reduction of the impact on passengers turns out to be a worsening of the weighted maintenance completion time with respect to the priority-based strategy where, on the contrary, only this parameter is optimized. This increase is instead smaller for the position-based strategy in which the priority order is not followed and the weighted completion time already presents a high value.

TABLE XII
Milan graph sizes.

| NODES | |
|---|---|
| Number of metro nodes | 19 |
| Number of alternative service nodes | 23 |
| *Total number of nodes* | *42* |
| LINKS | |
| Number of metro links | 36 |
| Number of alternative service links | 30 |
| Number of pedestrian links | 56 |
| Number of additional links | 56 |
| *Total number of links* | *178* |

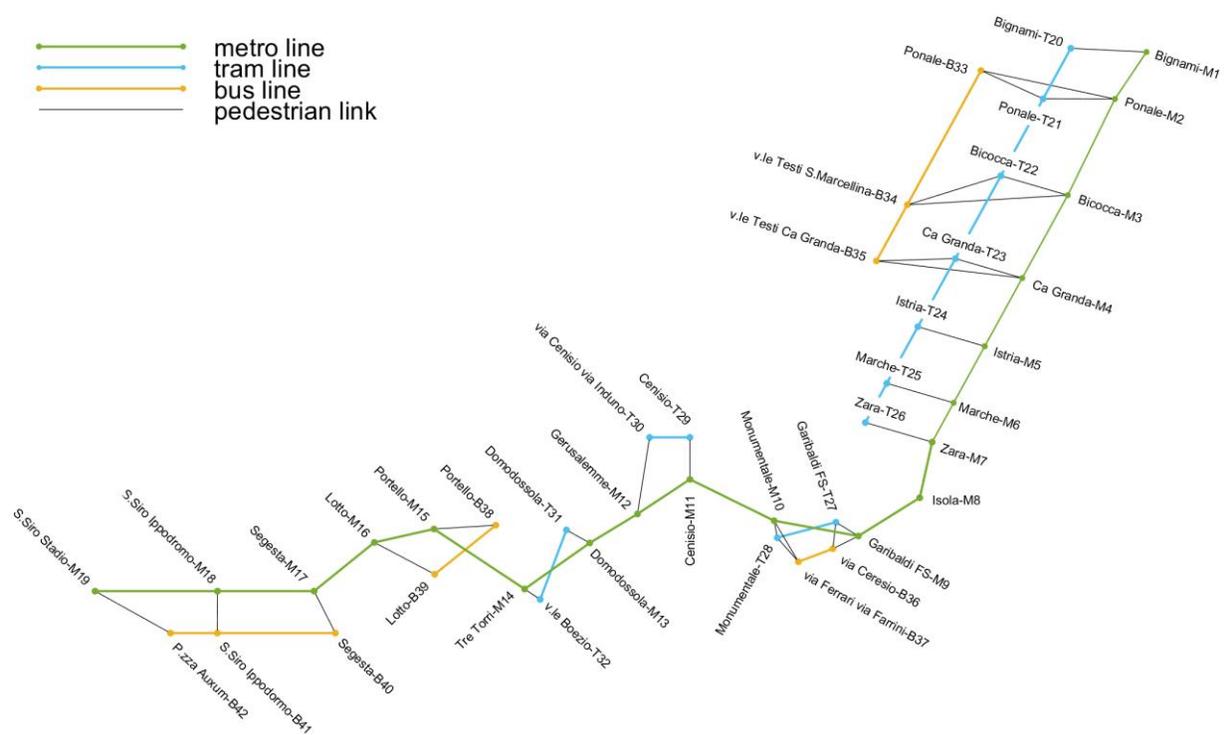

Fig. 11. The existing Milan multimodal network (without the activatable links).

TABLE XIII
Maintenance interventions.

| Link ID | Section | Duration (weeks) | Priority | Deadline (weeks) |
|---|---|---|---|---|
| 3 | Ponale - Bicocca | 1 | 9 | 10 |
| 9 | Istria - Marche | 1 | 2 | 20 |
| 13 | Zara - Isola | 1 | 1 | 20 |
| 15 | Isola - Garibaldi | 2 | 7 | 10 |
| 19 | Monumentale - Cenisio | 1 | 8 | 10 |
| 23 | Gerusalemme - Domodossola | 2 | 4 | 20 |
| 25 | Domodossola - Tre Torri | 1 | 6 | 20 |
| 27 | Tre Torri - Portello | 2 | 3 | 20 |
| 31 | Lotto - Segesta | 2 | 5 | 20 |

TABLE XIV
Computational time averaged over 15 randomly generated problem instances.

| | Average Solution Time (AST) [s] | Standard deviation [s] | Problem variables | Constraints |
|---|---|---|---|---|
| **SP1 (single instance)** | 0.37 | 0.06 | 922 | 27504 |
| **SP2 (single instance)** | 21.89 | 9.56 | 308859 | 381476 |
| **SP3 (single instance)** | 0.57 | 0.13 | 13825 | 14276 |
| **Overall problem - $R = 10$ iterations (each iteration implies solving SP2 and SP3 for 20 weeks and 3 o/d matrices)** | $13.13 \cdot 10^3$ | 0.24 | | |
| **Overall problem – $R = 20$ iterations (each iteration implies solving SP2 and SP3 for 20 weeks and 3 o/d matrices)** | $26.95 \cdot 10^3$ | 0.31 | | |

TABLE XV
Solution comparison with respect to the best solution of SP1 (averaged over 15 randomly generated problem instances and $R = 10$).

| Objectives | Variation with respect to the first feasible solution | Variation with respect to the position-based strategy | Variation with respect to the priority-based strategy |
|---|---|---|---|
| **Weighted maintenance completion time based on priority** | -52% | 7% | 38% |
| **Impact on passengers** | -26% | -30% | -30% |
| **Total** | -33% | -25% | -23% |

TABLE XVI
Solution comparison with respect to the best solution of SP2 (averaged over 15 randomly generated problem instances and $R = 10$).

| | Objectives | Variation with respect to the first feasible solution | Variation with respect to the position-based strategy | Variation with respect to the priority-based strategy |
|---|---|---|---|---|
| **Morning Peak** | Alternative service activation cost | -47% | -54% | -48% |
| | Pedestrian links generalized cost | -3% | -3% | 0% |
| | Metro links generalized cost | 6% | 11% | 9% |
| | Bus links generalized cost | 2% | 7% | 6% |
| | Train links generalized cost | -16% | -20% | -8% |
| | Additional links generalized cost | -51% | -56% | -54% |
| | Unsatisfied demand cost | -49% | -64% | -61% |
| **Evening Peak** | Alternative service activation cost | -46% | -56% | -53% |
| | Pedestrian links generalized cost | -3% | -3% | 1% |
| | Metro links generalized cost | 7% | 10% | 9% |
| | Bus links generalized cost | 1% | 5% | 4% |
| | Train links generalized cost | -18% | -25% | -10% |
| | Additional links generalized cost | -51% | -58% | -57% |
| | Unsatisfied demand cost | -46% | -57% | -56% |
| **Off Peak** | Alternative service activation cost | -36% | -47% | -50% |
| | Pedestrian links generalized cost | -7% | -8% | -2% |
| | Metro links generalized cost | 4% | 7% | 6% |
| | Bus links generalized cost | 8% | 18% | 14% |
| | Train links generalized cost | -26% | -36% | -13% |
| | Additional links generalized cost | -38% | -48% | -50% |
| | Unsatisfied demand cost | -46% | -59% | -61% |

22TABLE XVII
Solution comparison with respect to the best solution of SP3 (averaged over 15 randomly generated problem instances and $R = 10$).

|  | Objective Objectives | Variation with respect to the first feasible solution | Variation with respect to the position-based strategy | Variation with respect to the priority-based strategy |
|---|---|---|---|---|
| **Morning Peak** | Needed line capacity | -25% | -31% | -32% |
| | Needed links capacity | -36% | -33% | -28% |
| | Capacity surplus | -20% | -23% | -35% |
| **Evening Peak** | Needed line capacity | -21% | -28% | -37% |
| | Needed links capacity | -24% | -27% | -41% |
| | Capacity surplus | -15% | -25% | -30% |
| **Off Peak** | Needed line capacity | -17% | -26% | -27% |
| | Needed links capacity | -20% | -30% | -38% |
| | Capacity surplus | -22% | -23% | -34% |

Considering the quality of the solution of SP2, the comparison with the position and priority-based strategies (again considered as alternatives to the solution of SP1), reported in Table XVI, shows a significant reduction of the alternative service activation cost, of the unmet demand cost, and of the generalized cost relevant to the additional activated links. In contrast to the smaller case study, in this case, the average reduction of the unsatisfied demand cost is greater than zero for the off-peak demand due to a higher transport demand.

To conclude, SP3, reported in Table XVII, also has good performance with respect to the reference strategies.

*C. General comments on the results*

The presented case studies showed, from the point of view of the impact on the users, a very good performance in terms of improvement of the solution quality. In particular, the solutions found in the first iteration of the algorithm ($r = 1$) are generally better than those provided by the two reference strategies considered as alternatives to SP1. Such a behavior is due to the fact that the solution provided by SP1 in the first iteration already takes into account in the cost function the utilization rates $g_k, \forall k \in \mathcal{T}$, and, as a consequence, the first maintenance plan, assumed as input by SP2, already reduces the interaction between the maintenance activities and the users' demand. This effect is presented in Fig. 8 which shows that no maintenance activities are planned from the 6th week to the 10th one (i.e., week characterized by high demand rate), and during the 14th week (i.e., the one characterized by the demand peak).

In addition, the performances are further improved in the best solution, proving the positive effect of the proposed tabu search approach. In this connection, it is worth saying that the convergence capabilities of the proposed solution algorithm are good. In fact, in both case studies and in all the considered problem instances randomly generated, the best solution was found within the first 10 algorithm iterations. This is the reason why the detailed analysis of the quality of the solutions has been described only with reference to the setting $R = 10$.

Concerning the computational effort, it is easy to note that, given the structure of the solution algorithm, it results to be linear in the maximum number of iterations $R$, as confirmed in Table IX and XIV. In any case, the solution time is always acceptable for the kind of considered tactical problem, although an increase of the number of weeks and/or considering larger networks may result in a nonpractical solution time, especially of SP2, that would require an ad hoc solution algorithm, which is beyond the scope of this paper.

VII. CONCLUSIONS

The present paper describes a model that is able to support the negotiation process for booking maintenance possession windows and planning ad hoc alternatives during metro line maintenance operations. In more detail, the considered approach is an offline planning model which considers decisions to be made in advance to deal with planned preventive maintenance interventions that imply the unavailability of a section of the metro line for a time period of several days.

The proposed approach consists of a multistage cooperative framework with feedback that mimics the real-world decision-making process, dividing the global problem into three subproblems that represent different decision moments. The first subproblem considers the booking of rail possessions; the second subproblem tackles with the decision to use already existing services or activate ad hoc solutions; finally, the third subproblem addresses the design of the proposed alternative solution.

The solution approach is based on a negotiation algorithm and a tabu search strategy in which, for each solution of the first subproblem, the second and third subproblems are iteratively solved considering three different OD matrixes (i.e., those associated to morning peak, to evening peak, and to off-peak transport demand) and for a total time corresponding to several weeks.

To prove the general applicability of the model and test the computational effort, the proposed approach is applied to two case studies: the public transport networks of the Italian cities of Genoa and Milan. The two case studies are characterized by different network size and different transport demand. In particular, the graph of the Milan network is characterized by a larger number of nodes and links with respect to the Genoa case study.

The results show, for both the case studies, the possible advantages of a cooperative approach among different service providers in the case of planned service interruptions. The main benefits achieved are the cost reduction for service providers and the mitigation of the impact on users' travel experience in terms of reduction of the total unmet transport demand.

In particular, the proposed approach has also been proven to be a suitable methodology for dealing with demand fluctuations and forecasted high peaks. Moreover, the proposed problem decomposition allows the application of the three subproblems independently to deal with real-world situations when part of the decisions have already been made, when different constraints in the maintenance planning must be considered (i.e., SP1 is customized), or when different service providers have specific constraints in the determination of the bus lines (i.e., SP3 is customized). Finally, the proposed approach might be applied to case studies characterized by a different level of cooperation between service providers, such as public transport network in which the same provider controls different transport modes, or situations in which different service providers are willing to cooperate. The application of the proposed cooperative approach is particularly convenient in urban areas, where there are many alternative solutions, such as metro lines and public ground transport services (i.e., light trains and buses). In these cases, a set of specific subproblems SP3 can be customized for the different operators and solved in parallel.

Further developments of the study will consider the application of multimodal assignment models and stochastic users' behavior, removing the assumptions of uncongested network and deterministic choice. Moreover, heuristic approaches will be applied to reduce the computational complexity of the model.

## ACKNOWLEDGMENT

The authors wish to thank the Municipality of Genoa, which provided the demand data, and Mr. Zekarias Tewoldemedhin, who conducted part of the experiments of the second case study.

APPENDIX

The three o/d matrices considered in the first case study, for the morning peak, the evening peak, and the off-peak transport demand, are reported in Tab. XVIII, XIX and XX.

Moreover, the detailed solutions of SP1 and SP2 of the first case study are described here. In particular:

- In Table XXI, only the results for the morning and evening peaks are reported since no activation is suggested for the off-peak demand. Moreover, only the time intervals during which an additional service is suggested are reported. For the interruptions of links 3 during week 1, link 7 during weeks 3 and 4, and link 4 during weeks 15-16, the results show that no alternative links are activated.
- During week 14, the activation of additional links is suggested to satisfy the high transport demand expected towards the city center.
- During week 2, the additional service is activated only for the morning peak, whereas in weeks 11, 17, 18 and 19, the additional service is activated only for the evening peak.
- Fig. 12 shows that during the interruption of link 12 (Dinegro – Principe), in time intervals 12 and 13, two different lines are activated: Brin – Principe and Dinegro – Principe – Darsena.
- The required capacity is different for the morning and evening peaks: 800 and 1000 passengers/hour, respectively, for the morning peak demand and 1000 and 700 passengers/hour, respectively, for the evening peak demand. In this case, cooperation with the existing bus line Principe-Darsena is not suggested by the model result.
- Fig. 13 shows that during week 2, to cope with the interruption of metro link S. Giorgio – Sant'Agostino, the model result suggests cooperation and integration with the existing bus line De Ferrari – Darsena, with a capacity of 500 passengers/hour, only for the morning peak demand.
- In Fig. 14, the alternative services activated during week 14 are depicted. As mentioned, these services are needed, even if no interruption is planned in the time interval, to meet the high transport demand towards the city center due to the special event planned. The result shows that the capacity of an existing service is increased (link Principe – Darsena).

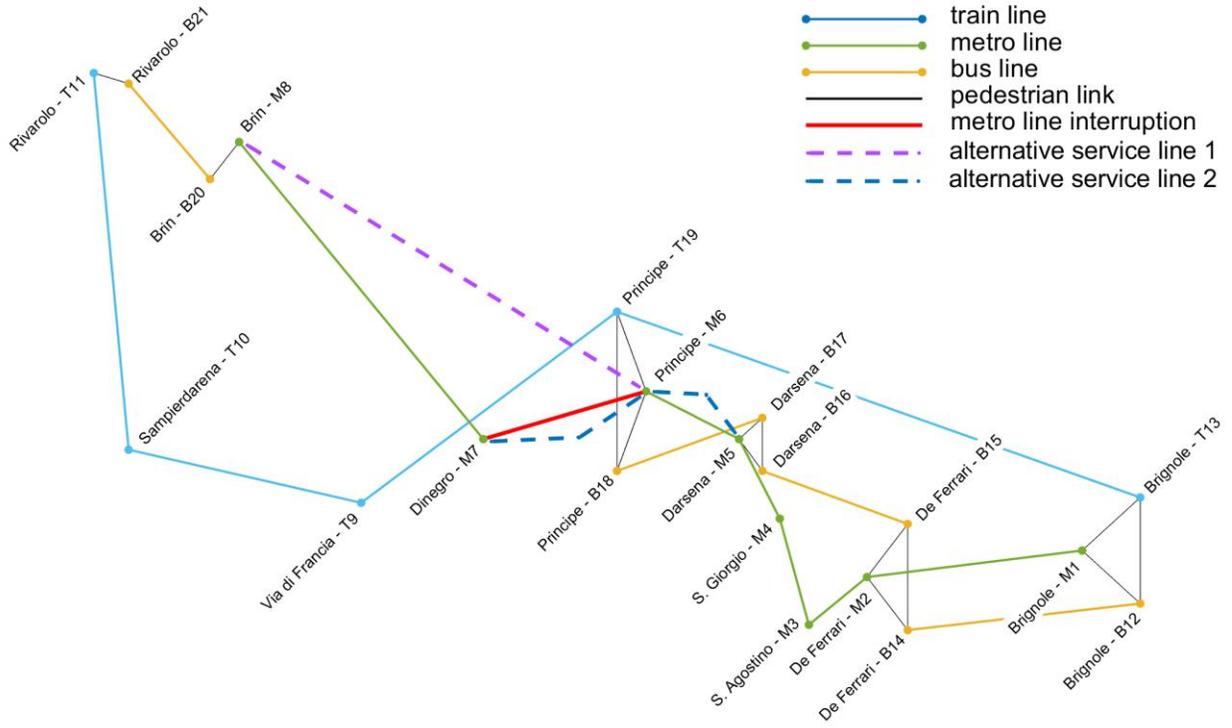

Fig. 12 Genoa case study: additional service during weeks 12-13 – Morning and Evening demand peak.

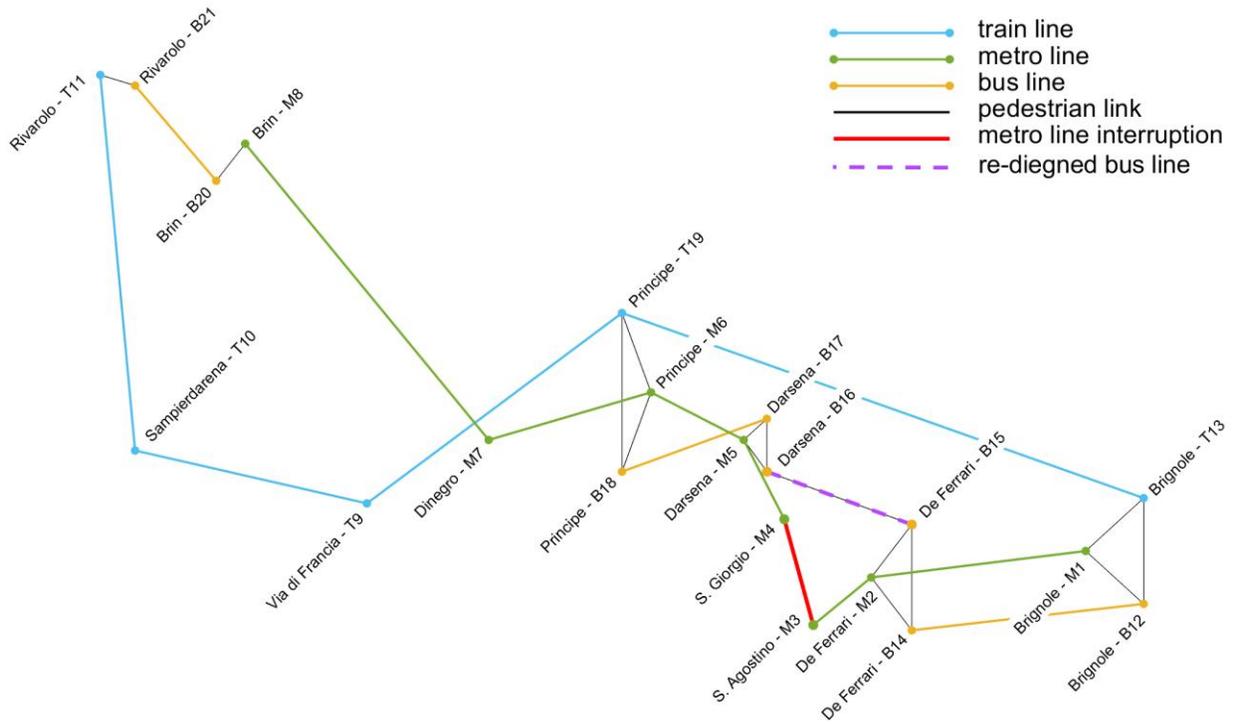

Fig. 13 Genoa case study: additional service during week 2 – morning peak.

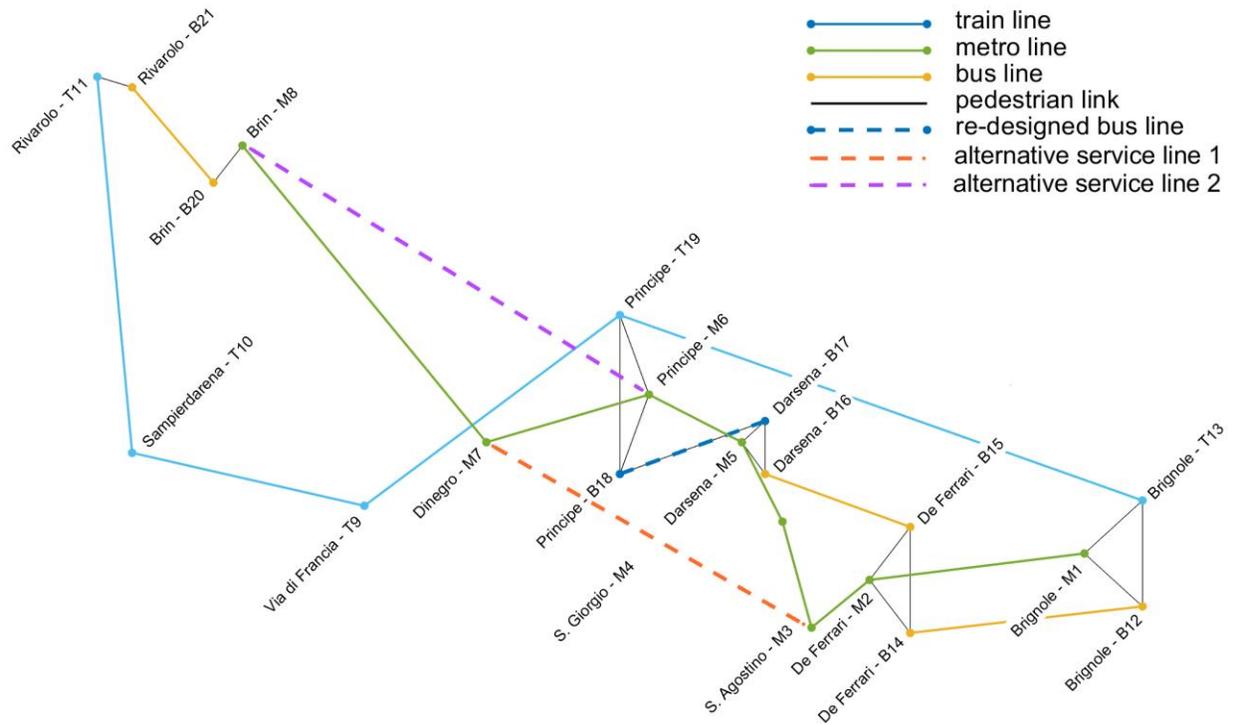

Fig. 14 Genoa case study: additional service during week 14 – demand peak due to special event.

TABLE XVIII
Morning peak o/d Matrix.

|     | 1M  | 2M  | 3M  | 4M  | 5M  | 6M | 7M | 8M | 9T | 10T | 11 T | 12B | 13T | 14B | 15B | 16B | 17B | 18B | 19T | 20B | 21B |
|-----|-----|-----|-----|-----|-----|----|----|----|----|-----|------|-----|-----|-----|-----|-----|-----|-----|-----|-----|-----|
| 1M  | 0   | 300 | 258 | 125 | 315 | 12 | 15 | 10 | 4  | 17  | 17   | 20  | 25  | 22  | 30  | 63  | 25  | 34  | 28  | 2   | 3   |
| 2M  | 250 | 0   | 179 | 130 | 320 | 14 | 4  | 8  | 15 | 12  | 22   | 25  | 26  | 21  | 34  | 34  | 31  | 21  | 35  | 3   | 9   |
| 3M  | 244 | 252 | 0   | 282 | 265 | 10 | 15 | 13 | 12 | 15  | 13   | 10  | 11  | 19  | 13  | 20  | 15  | 16  | 27  | 2   | 4   |
| 4M  | 273 | 231 | 214 | 0   | 224 | 6  | 13 | 11 | 16 | 2   | 15   | 10  | 14  | 20  | 16  | 13  | 26  | 24  | 28  | 3   | 5   |
| 5M  | 273 | 286 | 255 | 265 | 0   | 11 | 6  | 5  | 8  | 10  | 6    | 11  | 16  | 14  | 19  | 20  | 25  | 27  | 26  | 1   | 3   |
| 6M  | 241 | 283 | 257 | 280 | 264 | 0  | 4  | 2  | 4  | 6   | 14   | 12  | 8   | 16  | 22  | 26  | 18  | 19  | 18  | 1   | 4   |
| 7M  | 228 | 256 | 276 | 297 | 299 | 5  | 0  | 17 | 15 | 0   | 0    | 11  | 19  | 16  | 18  | 25  | 25  | 19  | 11  | 2   | 7   |
| 8M  | 243 | 256 | 218 | 236 | 282 | 6  | 7  | 0  | 7  | 2   | 3    | 0   | 12  | 14  | 6   | 13  | 18  | 14  | 17  | 1   | 3   |
| 9T  | 281 | 273 | 206 | 236 | 246 | 5  | 8  | 0  | 0  | 1   | 0    | 2   | 16  | 3   | 17  | 4   | 11  | 8   | 17  | 1   | 7   |
| 10T | 290 | 215 | 270 | 243 | 292 | 4  | 9  | 2  | 4  | 0   | 1    | 6   | 10  | 13  | 16  | 15  | 15  | 18  | 16  | 2   | 8   |
| 11T | 226 | 215 | 291 | 225 | 248 | 2  | 0  | 3  | 6  | 8   | 0    | 5   | 16  | 15  | 17  | 20  | 18  | 19  | 15  | 4   | 3   |
| 12B | 10  | 10  | 20  | 11  | 18  | 0  | 2  | 4  | 1  | 7   | 0    | 0   | 20  | 15  | 15  | 18  | 18  | 18  | 15  | 4   | 7   |
| I3T | 12  | 18  | 15  | 12  | 19  | 7  | 9  | 8  | 4  | 9   | 1    | 0   | 0   | 18  | 20  | 16  | 15  | 18  | 19  | 3   | 6   |
| 14B | 16  | 23  | 23  | 19  | 15  | 0  | 1  | 2  | 5  | 6   | 4    | 3   | 19  | 0   | 19  | 17  | 16  | 17  | 19  | 7   | 6   |
| 15B | 14  | 28  | 16  | 15  | 20  | 0  | 3  | 2  | 1  | 1   | 2    | 7   | 17  | 15  | 0   | 17  | 15  | 16  | 16  | 2   | 4   |
| 16B | 11  | 18  | 20  | 17  | 19  | 3  | 2  | 9  | 4  | 4   | 4    | 8   | 18  | 16  | 17  | 0   | 16  | 16  | 15  | 1   | 0   |
| 17B | 29  | 19  | 14  | 18  | 14  | 0  | 3  | 7  | 4  | 0   | 1    | 1   | 15  | 17  | 15  | 20  | 0   | 18  | 20  | 2   | 7   |
| 18B | 25  | 22  | 17  | 20  | 18  | 4  | 1  | 5  | 3  | 5   | 7    | 6   | 15  | 18  | 16  | 20  | 19  | 0   | 15  | 4   | 6   |
| 19 T| 15  | 14  | 29  | 14  | 16  | 7  | 4  | 3  | 6  | 3   | 4    | 1   | 17  | 16  | 15  | 20  | 16  | 20  | 0   | 1   | 7   |
| 20B | 24  | 26  | 28  | 28  | 27  | 0  | 2  | 8  | 9  | 2   | 8    | 4   | 16  | 18  | 15  | 19  | 15  | 18  | 18  | 0   | 0   |
| 21 B| 16  | 16  | 11  | 24  | 28  | 5  | 7  | 2  | 2  | 1   | 7    | 4   | 15  | 16  | 19  | 18  | 20  | 15  | 16  | 6   | 0   |

TABLE XIX
Evening peak o/d Matrix.

|     | 1M | 2M | 3M | 4M | 5M | 6M  | 7M  | 8M  | 9T  | 10T | 11 T | 12B | 13T | 14B | 15B | 16B | 17B | 18B | 19T | 20B | 21B |
|-----|----|----|----|----|----|-----|-----|-----|-----|-----|------|-----|-----|-----|-----|-----|-----|-----|-----|-----|-----|
| 1M  | 0  | 2  | 5  | 3  | 8  | 277 | 279 | 226 | 299 | 207 | 223  | 299 | 4   | 3   | 1   | 5   | 0   | 4   | 0   | 27  | 23  |
| 2M  | 12 | 0  | 13 | 2  | 12 | 203 | 263 | 227 | 286 | 232 | 221  | 298 | 4   | 0   | 0   | 1   | 1   | 0   | 1   | 26  | 26  |
| 3M  | 9  | 6  | 0  | 12 | 8  | 200 | 285 | 298 | 298 | 268 | 271  | 232 | 2   | 5   | 1   | 4   | 1   | 5   | 0   | 21  | 21  |
| 4M  | 0  | 7  | 7  | 0  | 16 | 225 | 283 | 244 | 238 | 226 | 244  | 247 | 2   | 5   | 5   | 1   | 1   | 4   | 2   | 30  | 25  |
| 5M  | 7  | 15 | 1  | 5  | 0  | 251 | 242 | 219 | 263 | 296 | 246  | 250 | 3   | 3   | 5   | 5   | 1   | 0   | 4   | 29  | 20  |
| 6M  | 4  | 19 | 18 | 0  | 13 | 0   | 223 | 203 | 298 | 282 | 224  | 237 | 5   | 5   | 1   | 2   | 1   | 0   | 1   | 23  | 29  |
| 7M  | 16 | 3  | 5  | 10 | 7  | 205 | 0   | 201 | 261 | 241 | 292  | 201 | 1   | 2   | 2   | 5   | 3   | 0   | 1   | 30  | 28  |
| 8M  | 5  | 0  | 16 | 8  | 4  | 289 | 295 | 0   | 209 | 268 | 257  | 293 | 5   | 4   | 4   | 3   | 1   | 1   | 3   | 26  | 23  |
| 9T  | 5  | 15 | 3  | 2  | 1  | 271 | 200 | 285 | 0   | 300 | 200  | 207 | 0   | 1   | 1   | 4   | 5   | 2   | 4   | 21  | 21  |
| 10T | 15 | 0  | 1  | 12 | 20 | 205 | 200 | 261 | 294 | 0   | 250  | 213 | 4   | 2   | 5   | 1   | 0   | 5   | 5   | 30  | 26  |
| 11T | 15 | 20 | 13 | 20 | 17 | 277 | 271 | 293 | 279 | 296 | 0    | 299 | 0   | 5   | 0   | 0   | 3   | 0   | 3   | 29  | 23  |
| 12B | 2  | 0  | 3  | 3  | 3  | 20  | 27  | 30  | 24  | 24  | 21   | 0   | 5   | 2   | 4   | 2   | 1   | 4   | 5   | 27  | 26  |
| I3T | 5  | 5  | 1  | 1  | 3  | 30  | 20  | 23  | 25  | 27  | 26   | 29  | 0   | 0   | 1   | 4   | 4   | 4   | 1   | 29  | 25  |
| 14B | 0  | 2  | 3  | 1  | 0  | 25  | 23  | 20  | 30  | 26  | 29   | 29  | 5   | 0   | 4   | 1   | 3   | 2   | 5   | 21  | 21  |
| 15B | 2  | 2  | 0  | 3  | 5  | 29  | 24  | 26  | 27  | 29  | 29   | 24  | 5   | 4   | 0   | 3   | 3   | 5   | 4   | 29  | 21  |
| **16B** | 5 | 0 | 3 | 4 | 4 | 21 | 23 | 22 | 20 | 21 | 30 | 23 | 5 | 3 | 2 | 0 | 5 | 5 | 3 | 23 | 24 |
| **17B** | 4 | 5 | 0 | 4 | 4 | 30 | 27 | 26 | 24 | 22 | 26 | 22 | 2 | 2 | 0 | 3 | 0 | 1 | 5 | 23 | 26 |
| **18B** | 0 | 4 | 3 | 1 | 5 | 28 | 22 | 25 | 26 | 21 | 20 | 23 | 4 | 2 | 5 | 4 | 0 | 0 | 1 | 29 | 30 |
| **19 T** | 3 | 0 | 2 | 4 | 2 | 24 | 23 | 26 | 21 | 27 | 22 | 24 | 1 | 5 | 4 | 4 | 1 | 0 | 0 | 24 | 29 |
| **20B** | 1 | 5 | 4 | 3 | 4 | 23 | 27 | 29 | 29 | 25 | 21 | 26 | 0 | 5 | 5 | 1 | 2 | 5 | 0 | 0 | 25 |
| **21 B** | 3 | 2 | 2 | 1 | 4 | 30 | 21 | 21 | 26 | 20 | 24 | 29 | 1 | 5 | 2 | 2 | 0 | 0 | 1 | 22 | 0 |

TABLE XX
Off peak o/d Matrix.

|  | 1M | 2M | 3M | 4M | 5M | 6M | 7M | 8M | 9T | 10T | 11 T | 12B | 13T | 14B | 15B | 16B | 17B | 18B | 19T | 20B | 21B |
|---|---|---|---|---|---|---|---|---|---|---|---|---|---|---|---|---|---|---|---|---|---|
| **1M** | 0 | 18 | 80 | 3 | 29 | 91 | 92 | 23 | 25 | 58 | 88 | 4 | 3 | 1 | 0 | 0 | 2 | 5 | 1 | 2 | 0 |
| **2M** | 74 | 0 | 13 | 2 | 54 | 88 | 55 | 55 | 18 | 66 | 32 | 1 | 2 | 2 | 2 | 0 | 3 | 2 | 5 | 2 | 2 |
| **3M** | 61 | 57 | 0 | 16 | 28 | 38 | 49 | 87 | 5 | 85 | 39 | 0 | 4 | 0 | 1 | 5 | 0 | 2 | 4 | 2 | 1 |
| **4M** | 17 | 22 | 94 | 0 | 98 | 20 | 52 | 61 | 28 | 90 | 36 | 3 | 0 | 1 | 0 | 2 | 2 | 0 | 0 | 1 | 0 |
| **5M** | 32 | 90 | 72 | 65 | 0 | 70 | 30 | 56 | 12 | 94 | 98 | 2 | 1 | 5 | 2 | 3 | 0 | 0 | 4 | 4 | 0 |
| **6M** | 39 | 56 | 65 | 12 | 3 | 0 | 91 | 76 | 47 | 73 | 39 | 5 | 3 | 4 | 2 | 5 | 4 | 4 | 0 | 2 | 1 |
| **7M** | 88 | 32 | 72 | 87 | 8 | 82 | 0 | 99 | 48 | 91 | 6 | 3 | 5 | 0 | 1 | 4 | 5 | 5 | 5 | 2 | 2 |
| **8M** | 92 | 9 | 0 | 14 | 14 | 37 | 9 | 0 | 31 | 41 | 74 | 1 | 1 | 4 | 4 | 3 | 2 | 1 | 3 | 3 | 3 |
| **9T** | 77 | 20 | 100 | 96 | 92 | 94 | 24 | 30 | 0 | 14 | 94 | 1 | 2 | 5 | 2 | 5 | 0 | 4 | 2 | 0 | 5 |
| **10T** | 47 | 44 | 13 | 49 | 92 | 1 | 45 | 68 | 20 | 0 | 27 | 3 | 5 | 5 | 4 | 4 | 5 | 4 | 1 | 3 | 0 |
| **11T** | 3 | 1 | 1 | 5 | 1 | 2 | 5 | 1 | 5 | 3 | 0 | 1 | 3 | 2 | 1 | 3 | 2 | 3 | 0 | 5 | 5 |
| **12B** | 0 | 4 | 0 | 2 | 1 | 0 | 1 | 5 | 2 | 2 | 0 | 0 | 0 | 5 | 3 | 3 | 0 | 2 | 4 | 4 | 1 |
| **13T** | 0 | 5 | 4 | 1 | 5 | 3 | 2 | 0 | 3 | 3 | 1 | 3 | 0 | 0 | 3 | 2 | 0 | 1 | 0 | 2 | 3 |
| **14B** | 3 | 4 | 5 | 4 | 5 | 0 | 2 | 2 | 0 | 3 | 1 | 5 | 0 | 0 | 5 | 1 | 3 | 1 | 4 | 2 | 0 |
| **15B** | 4 | 1 | 3 | 2 | 2 | 0 | 3 | 5 | 5 | 5 | 1 | 2 | 3 | 2 | 0 | 0 | 2 | 2 | 3 | 3 | 1 |
| **16B** | 3 | 0 | 5 | 0 | 1 | 0 | 1 | 5 | 2 | 5 | 5 | 3 | 1 | 5 | 5 | 0 | 5 | 2 | 3 | 0 | 3 |
| **17B** | 1 | 2 | 1 | 0 | 4 | 1 | 5 | 1 | 0 | 0 | 3 | 5 | 2 | 5 | 0 | 0 | 0 | 0 | 0 | 3 | 4 |
| **18B** | 2 | 4 | 5 | 3 | 2 | 3 | 0 | 3 | 4 | 1 | 5 | 0 | 0 | 5 | 4 | 2 | 0 | 0 | 1 | 5 | 2 |
| **19 T** | 4 | 5 | 3 | 0 | 2 | 0 | 1 | 5 | 4 | 2 | 0 | 1 | 0 | 1 | 4 | 0 | 1 | 5 | 0 | 3 | 4 |
| **20B** | 5 | 5 | 1 | 3 | 0 | 4 | 5 | 2 | 1 | 2 | 2 | 0 | 5 | 0 | 5 | 0 | 3 | 3 | 1 | 0 | 0 |
| **21 B** | 2 | 0 | 0 | 3 | 0 | 4 | 3 | 0 | 1 | 0 | 4 | 2 | 4 | 2 | 3 | 1 | 1 | 1 | 2 | 2 | 0 |

TABLE XXI
Results of subproblems 2 and 3

| Time interval (week) | Interrupted link | Morning peak | | Evening peak | |
|---|---|---|---|---|---|
| | | Redesigned links (Additional (A) or Existing €) | Capacity (Passengers/hour) | Redesigned links (Additional (A) or Existing €) | Capacity (Passengers/hour) |
| **1** | De Ferrari → Sant Agostino | - | - | - | - |
| **2** | Sant Agostino → San Giorgio | De Ferrari ↔ Darsena € | 500 | - | - |
| **3 – 4** | San Giorgio → Darsena | - | - | - | - |
| **11** | Darsena → Principe | - | - | Principe ↔ Darsena (A)<br>Darsena ↔ Brin (A) | 700<br>500 |
| **12-13** | Dinegro → Principe | Darsena ↔ Principe (A)<br>Principe ↔ Dinegro (A)<br>Brin ↔ Principe (A) | 1000<br>1000<br>800 | Darsena ↔ Principe (A)<br>Principe ↔Dinegro (A)<br>Brin ↔ Principe (A) | 700<br>700<br>1000 |
| **14** | - | Brin ↔ Principe (A)<br>Principe ↔ Darsena (E)<br>Dinegro ↔S. Agostino (A) | 900<br>500<br>700 | Brin ↔ Principe (A)<br>Principe ↔ Darsena (E)<br>Dinegro ↔S. Agostino (A) | 900<br>500<br>700 |
| **15-16** | Sant Agostino → De Ferrari | - | - | - | - |
| **17** | Brignole ↔De Ferrari | Brignole ↔ De Ferrari (A) | 1000 | Brignole ↔ De Ferrari (E) | 500 |
| **18** | De Ferrari → Brignole | - | - | Brignole ↔ De Ferrari (E) | 600 |
| **19** | Dinegro → Brin | - | - | Brin ↔ Principe (A) | 1000 |